\documentclass{iopjournal}
\expandafter\let\csname equation*\endcsname\relax 
\expandafter\let\csname endequation*\endcsname\relax
\usepackage{amsmath,amssymb,mathtools,bm}
\newcommand{\acknowledgments}{\ack}
\newcommand{\affiliation}{\affil}

\usepackage{graphicx}
\usepackage{hyperref}
\usepackage[utf8]{inputenc}
\usepackage[T1]{fontenc}
\usepackage[english]{babel}
\usepackage{lmodern}

\newcommand{\ket}[1]{\left\lvert{#1}\right\rangle}
\newcommand{\bra}[1]{\left\langle{#1}\right\rvert}
\newcommand{\op}[2]{\left\lvert{#1}\middle\rangle\!\middle\langle{#2}\right\rvert}
\newcommand{\intinf}{\int\!}

\begin{document}
\articletype{Paper}

\title{Higher-Order Adiabatic Elimination in Atom-Cavity Systems and Its Impact on Spin-Squeezing Generation}

\author{Stefano Giaccari$^{1,*}$\orcid{0000-0001-9467-2403}, Giulia Dellea$^2$, Marco G. Genoni$^{2,3,*}$\orcid{0000-0001-7270-4742} and Gianluca Bertaina$^{1,*}$\orcid{0000-0002-9440-4537}}

\affiliation{$^1$Istituto Nazionale di Ricerca Metrologica, Strada delle Cacce 91, I-10135 Torino, Italy}

\affiliation{$^2$Dipartimento di Fisica ``Aldo Pontremoli'', Universit\`a degli Studi di Milano, via Celoria 16, I-20133 Milano, Italy}

\affiliation{$^3$INFN, Sezione di Milano, I-20133 Milano, Italy}

\affiliation{$^*$ Authors to whom any correspondence should be addressed.}

\email{s.giaccari@inrim.it, marco.genoni@unimi.it, and g.bertaina@inrim.it}

\begin{abstract}
Spin-squeezed states are metrologically useful quantum states where entanglement allows for enhanced sensing with respect to the standard quantum limit. Key challenges include the efficient preparation of spin-squeezed states and the scalability of estimation precision with the number $N$ of probes. Recently, in the context of the generation of spin-squeezed states via coupling of three-level atoms to an optical cavity, it was shown that increasing the atom-cavity coupling can be detrimental to spin squeezing generation, an effect that is not captured by the standard second-order adiabatic cavity removal approximation. We describe adiabatic elimination techniques to derive an effective Lindblad master equation up to third order for the atomic degrees of freedom. Numerical simulations show that the spin squeezing scalability loss is correctly reproduced by the reduced open system dynamics, highlighting the role of higher-order contributions. Furthermore, we conjecture an extension beyond leading order of the adiabatic elimination technique to the case of conditional dynamics under quantum non-demolition continuous measurement and fast cavity loss, whose reliability is again confirmed by numerical simulation of the dynamics and the corresponding behavior of spin squeezing as a function of $N$.
\end{abstract}


\section{Introduction}
Quantum metrology~\cite{Giovannetti:2011chh,RevModPhys.89.035002,PerspectiveMultiPar,ReviewVictor} aims to improve the precision of parameter estimation by exploiting quantum resources. A typical protocol involves preparing a suitable probe state, allowing it to interact with the system of interest to encode the parameter, and performing a final measurement optimized to extract the encoded information. Techniques developed in this context have enabled progress in gravitational-wave detection~\cite{Schnabel:2010rha}, high-precision atomic clocks~\cite{2011NaPho5203K,PedrozoPenafiel2020,Robinson_Directcomparisontwo_2024}, and interferometry~\cite{RafalPO,Giovannetti:2011chh}.

Since probes of size $N$ described by separable states achieve at best the standard quantum limit (SQL) linear in $N$, quantum metrology relies on highly entangled states whose quantum Fisher information scales more favorably with $N$. Examples include GHZ states~\cite{PhysRevLett.82.1345}, Dicke states~\cite{PhysRev.93.99}, and, crucially, spin-squeezed states~\cite{WINELAND_1992,KITAGAWA_1993,Wineland_Squeezedatomicstates_1994,SpinSqueezingReview}. Their metrological performance is commonly captured by the squeezing parameter $\xi^2$, for which $\xi^{2}<1$ indicates an improvement over the SQL.

A standard approach to generate spin squeezing is to engineer effective nonlinear interactions among the atoms. A paradigmatic example is the one-axis twisting (OAT) Hamiltonian~\cite{KITAGAWA_1993}, which arises naturally in cavity-QED systems. When atoms interact dispersively with a driven cavity mode, the intracavity field mediates all-to-all interactions, giving rise to OAT-type dynamics and thus to spin-squeezed states~\cite{SSMITH_2010,Braverman_2019,Li_CollectiveSpinLightLightMediated_2022}.
In this regime, the minimum squeezing parameter scales as $\xi^2_{\rm m} \sim N^{-2/3}$, while the presence of cavity losses reduces the scaling to $\xi^2_{\rm m} \sim N^{-2/5}$ as shown in Ref.~\cite{Barberena:2023ham}.

A conceptually different but operationally related strategy exploits the information carried by photons leaking from the cavity. Continuous homodyne detection of the output field implements a quantum non-demolition (QND) measurement of a collective spin component, and the resulting measurement back-action conditionally squeezes the atomic state~\cite{Thomsen2002,Thomsen2002a,GEREMIA2003,MolmerMadsen2004,MADSEN_2004,Albarelli_Ultimatelimitsquantum_2017,ROSSI_2020,Binefa_2021,Caprotti_Analysisspinsqueezinggeneration_2024}.
This mechanism has been demonstrated in several experiments~\cite{Kuzmich_GenerationSpinSqueezing_2000,Appel_2009,SchleierSmith2010,Bohnet_2014,Cox2016,Hosten2016a,Huang2023,Serafin_NuclearSpinSqueezing_2021,MolmerNatPhys}.
In the dispersive regime, QND measurement can, in principle, generate Heisenberg-limited squeezing, $\xi^2_{\rm m} \sim N^{-1}$. If no continuous feedback is used, the ensemble-averaged squeezing scales instead as $\xi^2_{\rm m} \sim N^{-2/3}$~\cite{Caprotti_Analysisspinsqueezinggeneration_2024}, however, the full metrological precision can still reach the Heisenberg limit once the stochastic spin dynamics is properly accounted for in the estimation protocol~\cite{Bohnet_2014}. 
A comprehensive theoretical analysis of both methods, including scenarios where both protocols are simultaneously implemented and where also single-atom decoherence is taken into account, has recently been carried out in~\cite{Barberena:2023ham}. 

The analysis of such protocols relies on both solving the full atom–cavity dynamics numerically, typically using tools such as the QuTiP library~\cite{QUTIP_2012,QUTIP_2013}, and employing reduction techniques that restrict the description to the degrees of freedom (DOFs) directly relevant for spin squeezing. Such reductions are essential to obtain analytical insight, for instance to derive the optimal scalings of the squeezing parameters discussed above. In cavity-mediated schemes, this is commonly achieved through adiabatic elimination of the cavity mode.

For unitary dynamics the elimination can be carried out directly at the Hamiltonian level using established methods~\cite{SchriefferWolff,GAMEL_2010,JAMES_2011}. In the open-system setting, one may resort to brute-force Taylor expansions of the master equation~\cite{DOHERTY_1999} or to semiclassical approximations tailored to the cavity-QED regime~\cite{Barberena:2023ham}. More recently, however, systematic and rigorous frameworks for adiabatic elimination in open quantum systems have been developed~\cite{KesslerPRA2012,7798963,Azouit_2017,10886784}. The methods of Refs.~\cite{7798963,Azouit_2017,10886784} apply to composite systems consisting of a weakly coupled slow subsystem and a fast, strongly dissipative one. In this setting, the long-time dynamics is captured by an effective evolution on a reduced manifold obtained by discarding the rapidly relaxing DOFs.

It is, however, crucial to understand the limitations and regimes of validity of such elimination techniques. In the context of spin squeezing generation, particularly via QND measurement, it has been shown that departing from the bad-cavity limit can significantly degrade performance, notably affecting the achievable squeezing and its scaling with the number of atoms $N$~\cite{Caprotti_Analysisspinsqueezinggeneration_2024}.

The primary goal of this work is to apply adiabatic elimination to an atomic system dispersively coupled to a lossy cavity mode, going beyond the leading-order approximation and analyzing the resulting spin squeezing dynamics. The motivation for such a program stems from the results of Ref.~\cite{Caprotti_Analysisspinsqueezinggeneration_2024}, where it was found that the observed loss of scalability in the minimum spin-squeezing parameter achievable via the QND-measurement-based protocol is completely missed by the leading second order. Here we observe that this loss of scalability occurs also for the deterministic unconditional protocol where spin squeezing is generated via an effective interaction induced by a non-zero detuning between the driving laser and the cavity field. We find an explanation for the loss of scalability for large $N$ in the full dynamics for both protocols, by including third-order terms in the adiabatically reduced dynamics. This challenges the standard lore that in a bad-cavity regime the essential dynamics is already captured by second-order adiabatic elimination, while making the case for the necessity to consider higher orders. Therefore, we first extend the formalism introduced in Ref.~\cite{10886784} by deriving explicit expressions up to third order, and obtaining the corresponding Markovian master equation in Lindblad form for the atomic subsystem alone. This derivation led us to conjecture a form for the stochastic master equation describing the QND monitoring approach.
In both cases, we then compute the spin-squeezing parameter for states evolving under this effective dynamics, demonstrating how third-order corrections modify the behavior compared to second-order results and yield improved agreement with full numerical simulations that include the cavity mode explicitly.

The manuscript is structured to prioritize the physical insights and metrological conclusions in the opening sections, while reserving the formal derivation of the effective master equations for the latter part of the paper. In Sec.~\ref{sec:mainresults}, we introduce the physical model and summarize our core findings: the Markovian and stochastic master equations obtained through higher-order adiabatic elimination. Sec.~\ref{sec:Results} is dedicated to extensive numerical simulations, where we compare the full and reduced dynamics to demonstrate how third-order terms fundamentally alter the scaling of the spin-squeezing parameter. We recommend that readers interested primarily in the physical implications and the limits of spin-squeezing generation focus on these two sections. The formal mathematical framework is detailed in the subsequent sections. In Sec.~\ref{sec:AdiabaticElBipartite}, we review the adiabatic elimination method for bipartite systems~\cite{7798963,Azouit_2017,10886784} and provide a novel derivation of the third-order expansion terms. In Sec.~\ref{sec:AtomCavity}, we apply this general framework to our specific atom-cavity system, for both unconditional and conditional dynamics, to formally derive the master equations discussed in the previous sections. Finally, we offer concluding remarks and potential outlook in Sec.~\ref{sec:Conclusions}.

\section{Physical model and overview of our main results}\label{sec:mainresults}
We consider an ensemble of $N$ atoms that can be effectively described as two-level systems $\{\ket{\uparrow}, \ket{\downarrow}\}$ interacting with a single cavity mode with resonance frequency $\omega_c$ and linewidth $\kappa$. The cavity is driven by a laser of frequency $\omega_d$, with detuning $\delta = \omega_d - \omega_c$, amplitude $\beta$, and with a phase chosen so that the driving term takes the standard form $\beta(\hat{a}^\dagger + \hat{a})$, where $\hat{a}$ denotes the cavity annihilation operator in the rotating frame of the drive. Under these assumptions, the system Hamiltonian reads
\begin{equation}\label{eq:hamiltonian}
 \hat{H} = -\delta \hat{b}^\dagger \hat{b} 
 + g\,\hat{S}_z\!\left(\hat{b}^\dagger \hat{b} + \alpha \hat{b}^\dagger + \alpha^\ast \hat{b}\right),
\end{equation}
where $\hat{S}_z = \sum_{i=1}^{N} (\ket{\uparrow}_i\!\bra{\uparrow} - \ket{\downarrow}_i\!\bra{\downarrow})/2$ is the collective spin operator along the $z$ axis, and $\hat{b} = \hat{a} - \alpha$ denotes the annihilation operator of the displaced cavity mode, chosen so that $\langle \hat{b} \rangle = 0$ in the stationary state in the absence of atom–cavity coupling. The displacement parameter
\begin{equation}
 \alpha = \frac{-i\beta}{-i\delta + \kappa/2}
\end{equation}
corresponds to the steady-state amplitude of the driven cavity field in the uncoupled limit, giving an average photon number $\langle \hat{a}^\dagger \hat{a} \rangle = |\alpha|^2$. See Sec.~\ref{sec:AtomCavity} for more details on the derivation of this model.

The full atom–cavity dynamics, including photon loss from the cavity, is governed by the Lindblad master equation
\begin{align}
\label{eq:MEoriginal}
\frac{d}{dt}\rho &= \mathcal{L}(\rho) = -i \hat{H}^{\times}(\rho) + \kappa\,\mathcal{D}[\hat{b}](\rho),
\end{align}
where $\hat{H}^{\times}(\bullet) \coloneqq [\hat{H}, \bullet]$, and $\mathcal{D}[\hat{L}](\bullet) \coloneqq \hat{L} \bullet \hat{L}^\dagger - \frac{1}{2} (\hat{L}^\dagger \hat{L} \bullet + \bullet \hat{L}^\dagger \hat{L})$ denotes the dissipator associated with jump operator $\hat{L}$.

\subsection{Adiabatic elimination in the bad-cavity limit}
In the so-called \emph{bad-cavity limit} the cavity decay rate $\kappa$ is much larger than the atom–cavity coupling $g$, and one can perform an \emph{adiabatic elimination} of the cavity mode. Up to second-order in the coupling $g$, this leads to an effective master equation for the atomic density operator $\rho_s$ only~\cite{Caprotti_Analysisspinsqueezinggeneration_2024,Barberena:2023ham}:
\begin{align}
\label{eq:MEsecondordert}
\frac{d}{dt}\rho_s &= -i \chi [ \hat{S}_z^2, \rho_s] + \tilde{\kappa}\,\mathcal{D}[\hat{S}_z](\rho_s),
\end{align}
where the effective dispersive coupling and dephasing rates are
\begin{align}\label{eq:chikappatilde}
 \chi &= \delta\,\frac{4 g^2 |\alpha |^2}{\kappa^2 + 4 \delta^2}, 
 &\tilde{\kappa} &= \frac{4 g^2 |\alpha|^2 \kappa}{\kappa^2 + 4 \delta^2}.
\end{align}
For nonzero detuning $\delta$, this corresponds to an effective OAT Hamiltonian, which can generate spin-squeezed states~\cite{KITAGAWA_1993}.

\subsection{Conditional dynamics under continuous measurement}

The above dissipative dynamics can be extended by introducing continuous homodyne monitoring of the cavity output with efficiency $\eta \in [0,1]$. The conditional evolution of the system is then described by the stochastic master equation (SME) in It\^{o} form~\cite{Wiseman_Quantumtheoryfieldquadrature_1993,Wiseman_QuantumMeasurementControl_2009,Albarelli2024}:
\begin{align}
 d\rho &= \mathcal{L}(\rho)\,dt 
 + \sqrt{\eta}\,\mathcal{H}[\sqrt{\kappa}\,\hat{b} e^{-i\phi}](\rho)\,dW_t, \label{eq:StochasticFullMasterEq}\\
 I(t)\,dt &= \sqrt{\eta\kappa}\,
 \langle \hat{b} e^{-i\phi} + \hat{b}^\dagger e^{i\phi} \rangle dt + dW_t, \label{eq:signal}
\end{align}
where $\mathcal{H}[\hat{L}](\bullet) \coloneqq \hat{L} \bullet + \bullet \hat{L}^\dagger - \langle \hat{L}^\dagger + \hat{L}\rangle \bullet$ is the measurement back-action superoperator, $\phi$ is the homodyne phase selecting the measured quadrature, and $dW_t$ is a Wiener increment. The averages $\langle \cdot \rangle$ are taken over the conditional state $\rho(t)$, which introduces a nonlinearity in the SME.

Adiabatic elimination at the second order has also been applied in this conditional setting, yielding the effective SME~\cite{DOHERTY_1999,Barberena:2023ham}
\begin{align}
d\rho_s &= -i \chi [\hat{S}_z^2, \rho_s]\,dt 
+ \tilde{\kappa}\,\mathcal{D}[\hat{S}_z](\rho_s)\,dt  + \sqrt{\eta\tilde{\kappa}}\,\mathcal{H}[\hat{S}_z e^{-i(\phi-\theta_s)}](\rho_s)\,dW_t, 
\label{eq:Stochastic_2Order}\\
I(t)\,dt &= 2\cos(\phi-\theta_s)\sqrt{\eta\tilde{\kappa}}\,\langle \hat{S}_z\rangle dt + dW_t,
\label{eq:signal_2Order}
\end{align}
where the reduced jump phase
\begin{equation}\label{eq:JumpOverallPhase}
 e^{i\theta_s} = \frac{(d - i)^2}{1 + d^2},
\end{equation}
with $d = 2\delta/\kappa$, does not affect the unconditional dissipator but plays a role in the conditional dynamics. In this setting, the regime of interest for spin squeezing corresponds to zero detuning ($\delta=0$), where an appropriate choice of the homodyne phase $\phi$ realizes a continuous QND measurement of $\hat{S}_z$, resulting in spin squeezing along this direction~\cite{Thomsen2002a,Thomsen2002,Caprotti_Analysisspinsqueezinggeneration_2024}.

However, as already pointed out in Ref.~\cite{Caprotti_Analysisspinsqueezinggeneration_2024} and discussed in the following, the above second-order effective models can significantly overestimate the degree of spin squeezing compared to the full atom–cavity dynamics. In particular, they may yield an incorrect scaling of the achievable squeezing with the total number of atoms $N$.

\subsection{Main result: third-order effective master equations}

This subsection presents the central theoretical result of this work: 
the derivation of a third-order effective master equation for the atomic subsystem, 
obtained via a systematic adiabatic elimination of the cavity mode. 
This higher-order treatment captures dynamical effects beyond the standard second-order approximation 
and is essential to correctly reproduce the scaling of spin squeezing with the number of atoms~$N$.

Building on recently developed systematic adiabatic elimination techniques for open quantum systems, valid to arbitrary order in the perturbative parameter $\epsilon = g/\kappa$, we derive here the effective atomic master equation up to third order:
\begin{align}
\label{eq:MEtotalPositive}
 \frac{d}{dt}\rho_s 
 &= \mathcal{L}_s(\rho_s) 
 = -i \hat{H}_s^{\times}(\rho_s) + \mathcal{D}[\hat{L}_s](\rho_s),
\end{align}
with the third-order effective Hamiltonian
\begin{equation}\label{eq:Hamiltonian3order}
 \hat{H}_s = 
 \frac{4 n_0 \kappa}{(1 + d^2)^2}
 \left(
 \frac{d}{2}\epsilon^2 \hat{S}_z^2
 - \frac{1 - d^2}{1 + d^2}\epsilon^3 \hat{S}_z^3
 \right),
\end{equation}
where $n_0 = (2\beta/\kappa)^2$ is the stationary photon number of the uncoupled resonant cavity, and $|\alpha|^2 = n_0/(1 + d^2)$. The corresponding jump operator, accurate up to second order, is
\begin{equation}
\label{eq:Jump3order}
 \hat{L}_s = \sqrt{\tilde{\kappa}}\,e^{i\theta_s}
 \left(\hat{S}_z + \frac{2(d - i)\epsilon}{1 + d^2}\,\hat{S}_z^2\right).
\end{equation}
The detailed derivation of Eqs.~\eqref{eq:Hamiltonian3order}–\eqref{eq:Jump3order} is presented in 
Secs.~\ref{sec:AdiabaticElBipartite} and~\ref{sec:AtomCavity}. 
It is straightforward to verify that the second-order contribution to $\hat{H}_s$ 
and the leading term in $\hat{L}_s$ coincide with those appearing in the standard 
second-order effective master equation~\eqref{eq:MEsecondordert}.

\noindent
Starting from the reduced Lindbladian $\mathcal{L}_s$ in Eq.~\eqref{eq:MEtotalPositive}, 
we further propose that the conditional dynamics under continuous homodyne monitoring 
of the cavity output is described by the effective stochastic master equation
\begin{align}
 d\rho_{s} &= \mathcal{L}_s (\rho_{s})\,dt 
 + \sqrt{\eta}\,\mathcal{H}[\hat{L}_s e^{-i\phi}](\rho_{s})\,dW_t, 
 \label{eq:StochasticRedMasterEq}\\
 I(t)\,dt &= \sqrt{\eta}\,
 \langle \hat{L}_s e^{-i\phi} + \hat{L}_s^\dagger e^{i\phi}\rangle\,dt + dW_t, 
 \label{eq:signalRed}
\end{align}
which takes the standard form for a homodyne measurement of the reduced operator $\hat{L}_s$. 
This conjecture is the natural extension of the second-order SME~\eqref{eq:Stochastic_2Order} and we discuss its validity in Sec.~\ref{sec:AtomCavity}.

\medskip
In the following section, we assess the physical implications of the dynamics described by the effective master equations~\eqref{eq:MEtotalPositive} and~\eqref{eq:StochasticRedMasterEq}, and analyze the generated spin-squeezing. 
In particular, we compare the achievable squeezing and its scaling with the number of atoms~$N$ as predicted by the third-order effective dynamics, by the standard second-order approximation, and by the full atom--cavity master equation.
This comparison highlights the crucial role of higher-order corrections in accurately capturing the collective quantum correlations responsible for metrological enhancement.

\section{Numerical results for the spin-squeezing parameter}
\label{sec:Results}
We now consider how the dynamics described above can evolve an initially separable state into a spin-squeezed state. Having introduced the collective spin operator $\mathbf{\hat{S}}=\sum_{j=1}^{N}(\frac{1}{2}(\ket{\uparrow}_j\!\bra{\downarrow} + \ket{\downarrow}_j\!\bra{\uparrow}),\frac{1}{2i}(\ket{\uparrow}_j\!\bra{\downarrow} - \ket{\downarrow}_j\!\bra{\uparrow}),\frac{1}{2}(\ket{\uparrow}_j\!\bra{\uparrow} -\ket{\downarrow}_j\!\bra{\downarrow}))$, we initialize the system of $N$ qubits in a coherent spin state (CSS) along $+X$ $\ket{ \mathrm{CSS}_x} = \bigotimes_j^N \ket{\rightarrow_x}_j $, which is in the total spin sector $S=S_\mathrm{max}=N/2$. The evolution governed by the MEs and SMEs presented in the previous section remains in the $S_\mathrm{max}$ sector, and it produces squeezing of the collective spin, quantified by the spin-squeezing parameter~\cite{WINELAND_1992,Wineland_Squeezedatomicstates_1994}
\begin{equation}
 \xi^2 = N\frac{{\min}_\perp \text{Var}(\hat{S}^\perp)}{\vert\langle\mathbf{\hat{S}}\rangle\vert^2}\,,
\end{equation}
where ${\min}_\perp [\text{Var}(\hat{S}^\perp)]$ indicates the smaller of the two eigenvalues of the covariance matrix $\text{Var}(\hat{S}^\perp)$ of the collective spin components transverse to the polarization axis $\langle\mathbf{\hat{S}}\rangle/|\langle\mathbf{\hat{S}}\rangle|$. The initial CSS has $\xi^2 = 1 $, whereas $\xi^2 < 1 $ indicates squeezing of the uncertainty along one direction in the transverse plane with respect to the CSS state. 

Our goal is to compare full dynamics simulations with the reduced-dynamics ones obtained first by only considering second-order adiabatic elimination known in the literature and then by including the novel third-order contributions. Since we rigorously derived the unconditional master equation~\eqref{eq:MEtotalPositive} and argue for the validity of the conditional master equation~\eqref{eq:StochasticRedMasterEq}, we are in the position of investigating three regimes of interest: i) unconditional dynamics only, dominated by the effective OAT Hamiltonian, ii) QND dynamics with minimized Hamiltonian effect, namely at zero detuning, and iii) the simultaneous action of the two, at finite detuning. This case has been systematically investigated at the second-order level in Ref.~\cite{Barberena:2023ham}. Here, we focus on the role of beyond-leading-order effects. 
In all these scenarios we will present our numerical findings obtained by simulating both conditional and unconditional dynamics with the QuTiP library~\cite{QUTIP_2012, QUTIP_2013,Caprotti_Analysisspinsqueezinggeneration_2024}. The typical time evolution of the spin-squeezing parameter is exemplified in App.~\ref{app:data}, where details of the simulations are also provided. Starting from $\xi^2=1$, Hamiltonian evolution and/or continuous measurement bring the parameter to values much smaller than 1, depending on $N$. Then, the unavoidable loss of contrast due to curvature or dissipation effects brings again $\xi^2$ towards larger values. This evolution thus manifests an optimal time at which the minimum value of $\xi^2$ is reached, defining the optimal parameter $\xi^2_m$. In the main text, we collect such optimal $\xi^2_m$ for increasing $N$ and different parametric regimes. The numerical data produced in this study are available online~\cite{zenodoGiaccari2025}.
\subsection{Unconditional dynamics dominated by the OAT Hamiltonian}
\begin{figure*}[tbp]
 \centering
 \includegraphics[width=0.98\textwidth]{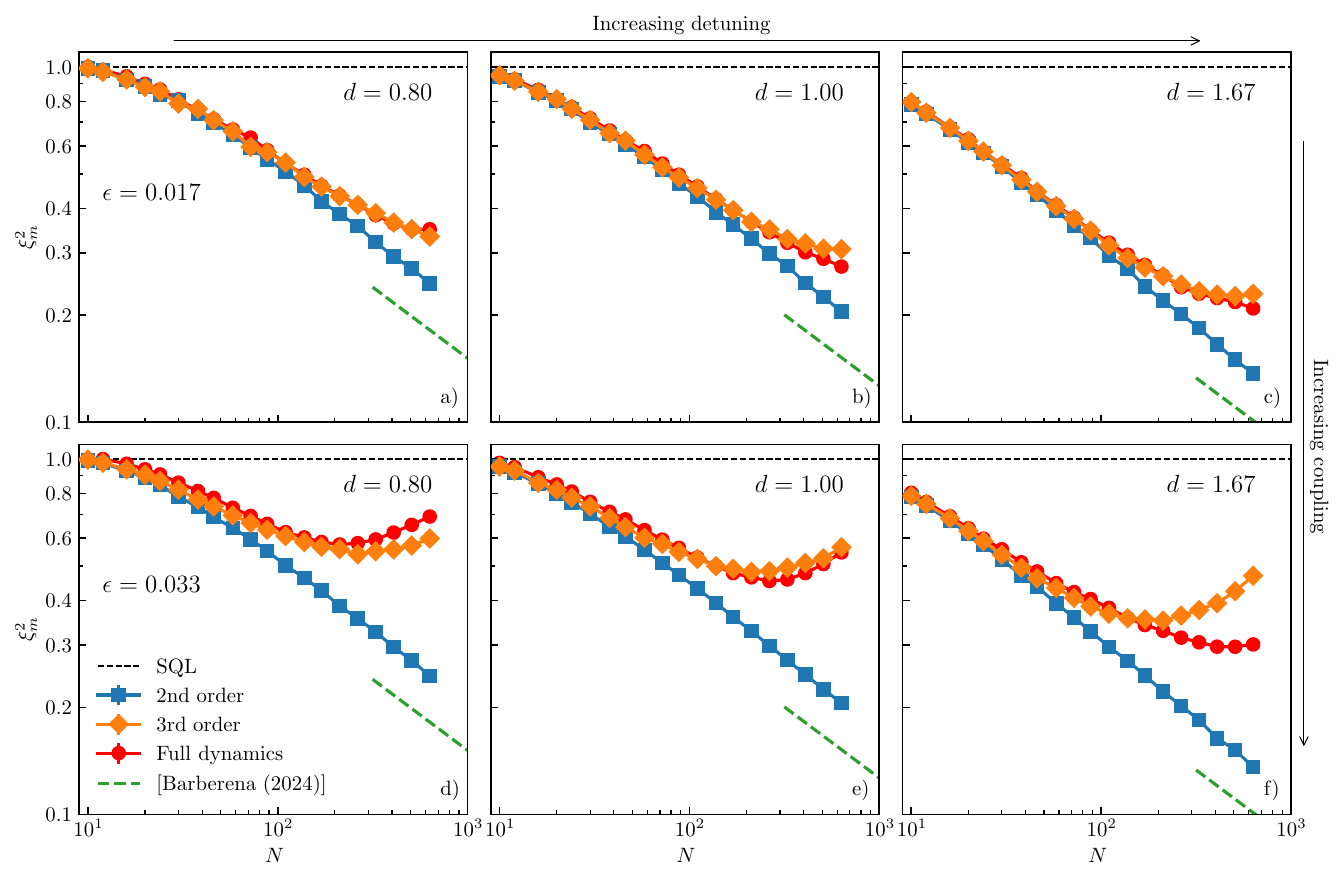}
 \caption{Values of the optimal spin-squeezing parameter with increasing number of atoms $N$, for coupling parameters $\epsilon=g/\kappa = 0.017$ (upper row) and $0.033$ (lower row), and driving laser detunings $d=2\delta/\kappa=0.8$ (first column), $1.0$ (second column), and $1.67$ (third column). Circles: solution of the full unconditional master equation~\eqref{eq:MEoriginal}. Squares: solution of the second-order master equation from Eq.~\eqref{eq:MEsecondordert}. Diamonds: solution of the third-order master equation~\eqref{eq:MEtotalPositive}. Dashed line: second-order asymptotic result, Eq.~\eqref{eq:scalingbarberena}. Horizontal dashed line: standard quantum limit. Statistical error bars are smaller than symbol sizes.}
 \label{fig:compare_unconditional}
\end{figure*}
We vary the number of atoms $N$ and study the quench dynamics starting from the spin-coherent state along the $+X$ direction, comparing to full evolutions starting from the same spin state and an empty cavity, varying $\epsilon$ and $d$ while fixing $\beta/\kappa=1$. We compare the predictions of the spin-squeezing parameters for increasing $N$ from Eq.~\eqref{eq:MEtotalPositive} against both the full master equation in Eq.~\eqref{eq:MEoriginal} and the second-order approximation from Eq.~\eqref{eq:MEsecondordert}. The latter portrays spin squeezing generation in a dissipative OAT model, in the absence of continuous measurement. Analytical predictions for the minimum spin-squeezing parameter $\xi^2_m$ in the large-$N$ limit, derived in Ref.~\cite{Barberena:2023ham} for different regimes of $\epsilon$ and $d$, also provide a useful benchmark. 
Under our assumptions, for the second-order ME~\eqref{eq:MEsecondordert} the spin-squeezing parameter is expected to scale as~\cite{Barberena:2023ham}:
\begin{equation}\label{eq:scalingbarberena}
\xi^2_m \simeq \frac{5}{2 (3 d^{4})^{1/5}}\frac{1}{N^{2/5}}\,,
\end{equation}
and we include this scaling behavior as a reference in our numerical comparisons. 

In Fig.~\ref{fig:compare_unconditional}, we show the results for values of $\epsilon = 0.017$ (first row), and $\epsilon = 0.033$ (second row) and $d=0.8$ (first column), $d=1.0$ (second column) and $d=1.67$ (third column). We immediately observe that the third-order master equation reproduces much more accurately the full atom-cavity dynamics, in particular reproducing the fact that the scaling $N^{-2/5}$, promised by the second-order adiabatic elimination, is lost for $N\approx 10^2$ already for values of $\epsilon\approx 2\cdot 10^{-2}$, that one could consider small enough to allow for a second-order expansion. Remarkably, in panels a-c, we find that for such a value of $\epsilon$
the third-order master equation is particularly accurate even for large $N>1/\epsilon$. This might seem surprising, because, by inspecting the interaction Hamiltonian, one would expect that the perturbative expansion accuracy is determined by $\epsilon\langle\hat{S}_z\rangle\sim \epsilon N$. However, we argue that the initial state plays an important role, since for the CSS state along $+X$ one has $\langle\hat{S}_z\rangle=0$ and $\langle\hat{S}_z^2\rangle=N/4$. We are led to infer that the relevant small parameter determining the degree of convergence of this reduced quench dynamics is $\epsilon\sqrt{\langle\hat{S}_z^2\rangle}\sim \epsilon \sqrt{N}$, while $N\sim 1/\epsilon$ corresponds to the order of magnitude of the optimal number of atoms for spin squeezing generation. 
We generally observe that with decreasing $\epsilon$ and increasing $d$ more spin squeezing is generated. This effect is not obvious, because it stems from both the Hamiltonian and dissipative parts of the dynamics. Recall, for example, that the purely Hamiltonian second-order result $\xi_m^2=3^{2/3}/2N^{2/3}$ is independent of $\epsilon$ and $d$, which only set the timescale of OAT dynamics. Effectively, Eqs.~\eqref{eq:Hamiltonian3order} and \eqref{eq:Jump3order} are an expansion in the parameter $\epsilon/\sqrt{1+d^2}$, which is largest for $d\to 0$. Since we keep $\epsilon$ fixed in each row of Fig.~\ref{fig:compare_unconditional}, beyond-leading-order effects are stronger in panel d.
In panel f, where $\epsilon=0.033$ and $d=1.67$, the third-order results, although more accurate than the second-order ones at small $N$, depart significantly from the full simulations for large $N$, by predicting less spin squeezing than is seen in the full simulations. 
This result is at variance with panel e, which corresponds to the same $\epsilon$ but for detuning $d=1$. This hints at a role of $d$ in determining the degree of convergence of the perturbative expansion in $\epsilon$. 
A possible explanation is the following: in the full simulations, we observed that, at fixed $\epsilon$ and with increasing $d>1$, the number of photons in the full dynamics fluctuates significantly before settling to the stationary value, analogously to the uncoupled case of a driven-dissipative cavity. 
Such large fluctuations in the number of photons arguably already introduce non-trivial squeezing dynamics, therefore it is not true that the reduced atomic density matrix $\text{Tr}_B(\rho)$ is still close to the initial CSS after the transient. 
While such transient mixing time always occurs, since $\kappa$ is finite, when $d$ is small the average number of photons only smoothly increases from zero to the stationary value, reducing its impact in transient squeezing generation. 
This explains the observation in panel f, that for increasing $\epsilon$ and large $d>1$ the third-order expansion is not as accurate as for $d\sim 1$. Notice also that for exactly $d=1$ the third-order Hamiltonian contribution in Eq.~\eqref{eq:Hamiltonian3order} vanishes. This implies that the significant departure of the full results from the second-order results in panels b and e is essentially due to the beyond-second-order correction of the dissipative contribution alone.
\subsection{QND conditional dynamics at zero detuning}
We now consider the scenario where the cavity output is continuously monitored via homodyne detection. If we further assume a resonant laser driving (zero detuning $d=0$) and fix the physical homodyne phase to $\phi=0$, we obtain an effective QND measurement of the $\hat{S}_z$ spin component. It is thus expected that each conditional state will exhibit measurement-induced spin squeezing. In this scenario we will consider as a figure of merit the spin-squeezing parameter averaged over all the possible conditional states. Numerically, this is obtained by performing a Monte Carlo sampling of the quantum trajectories as described in~\cite{Caprotti_Analysisspinsqueezinggeneration_2024}. As in the previous scenario, also here the (averaged) spin-squeezing parameter shows a non-monotonic behavior in time and we focus on the optimal (minimum) value $\xi_m^2$. We estimate also its Monte Carlo uncertainty, which is nevertheless not noticeable in the figures.

\begin{figure*}[tbp]
 \centering
 \includegraphics[width=0.80\textwidth]{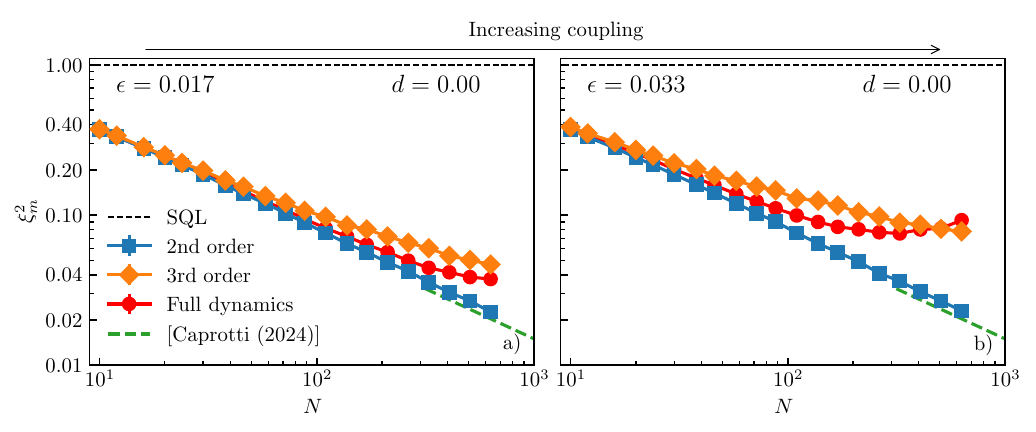}
 \caption{Values of the optimal (average) spin-squeezing parameter with increasing number of atoms $N$, for the conditional dynamics described by Eq.~\eqref{eq:StochasticRedMasterEq} with coupling parameters $\epsilon = 0.017$ (panel a), and $\epsilon = 0.033$ (panel b), for driving laser detuning $d=0$, homodyne phase $\phi=0$, and efficiency $\eta=1$. Circles: solution of the full Eq.~\eqref{eq:StochasticFullMasterEq}. Diamonds: solution of the third-order Eq.~\eqref{eq:StochasticRedMasterEq}. Squares: solution of Eq.~\eqref{eq:StochasticRedMasterEq} truncated to second order. Dashed line: second-order asymptotic result in the case $d=0$, Eq.~\eqref{eq:scalingcaprotti}. Horizontal dashed line: standard quantum limit. Statistical error bars are smaller than symbol sizes.}
 \label{fig:compare_conditional_d0}
\end{figure*}

In Fig.~\ref{fig:compare_conditional_d0}, we show the results of QuTiP simulations of the conditional dynamics with maximum efficiency $\eta=1$, given by the stochastic master equation~\eqref{eq:StochasticFullMasterEq} in the full Hilbert space, compared to the results from the third-order reduced master equation~\eqref{eq:StochasticRedMasterEq} and its truncation to second order. We simulate the couplings $\epsilon=0.017$ (panel a) and $\epsilon=0.033$ (panel b), as in Fig.~\ref{fig:compare_unconditional}. As a reference, we also plot the analytical second-order asymptotic result found in Ref.~\cite{Caprotti_Analysisspinsqueezinggeneration_2024} for the case $d=0$:
\begin{equation}\label{eq:scalingcaprotti}
\xi^2_m \simeq \frac{3}{2}\frac{1}{N^{2/3}}\,.
\end{equation}
We observe that this dynamics manifests more spin squeezing than the unconditional case of Fig.~\ref{fig:compare_unconditional}. It is indeed known that QND spin squeezing generation is favored in the resonant case (see Refs.~\cite{Li_CollectiveSpinLightLightMediated_2022,Barberena:2023ham}). The full conditional simulations show a minimum of the optimal spin-squeezing parameter in the considered range of $N$ for the larger $\epsilon$ coupling of panel b, an effect that was observed in Ref.~\cite{Caprotti_Analysisspinsqueezinggeneration_2024} and which is completely missed by the second-order conditional dynamics. The third-order result starts to manifest such a feature only at much higher $N$. As we discussed previously, this could be partially due to the effectively larger coupling when $d=0$.

The third-order result is overall more accurate than the second-order one, especially for large $N$. Whereas for sufficiently small $N$ third-order corrections are negligible, for intermediate $N$ they appear to give a less favorable minimum squeezing parameter than the one given by the full dynamics, which is instead better approximated by the second-order description. We rationalize this counterintuitive behavior by observing that our conditional dynamics is simulated with the standard form of the stochastic master equation for continuously monitored quantum systems, Eq.~\eqref{eq:StochasticRedMasterEq}, where jump operators in dissipation channels are equal to the measurement ones. This in particular means that the back-action terms in the second-order (third-order) SMEs contain the jump operators at first-order (second-order), insofar as these are the ones showing up in the dissipators. As commonly done in the literature~\cite{DOHERTY_1999,Barberena:2023ham}, we are therefore missing the contribution to the stochastic term that comes from the reduced measured operator calculated at next order. Arguably, including a $\hat{S}_z^3$ correction to the back-action term in the third-order SME, would improve the spin-squeezing parameter, and, conversely, including the $\hat{S}_z^2$ correction to the back-action term in the second-order SME would make it worse, thus enhancing the accuracy of both approximations. We plan to numerically investigate this possibility in future work. 

\begin{figure*}[tbp]
 \centering
 \includegraphics[width=0.98\textwidth]{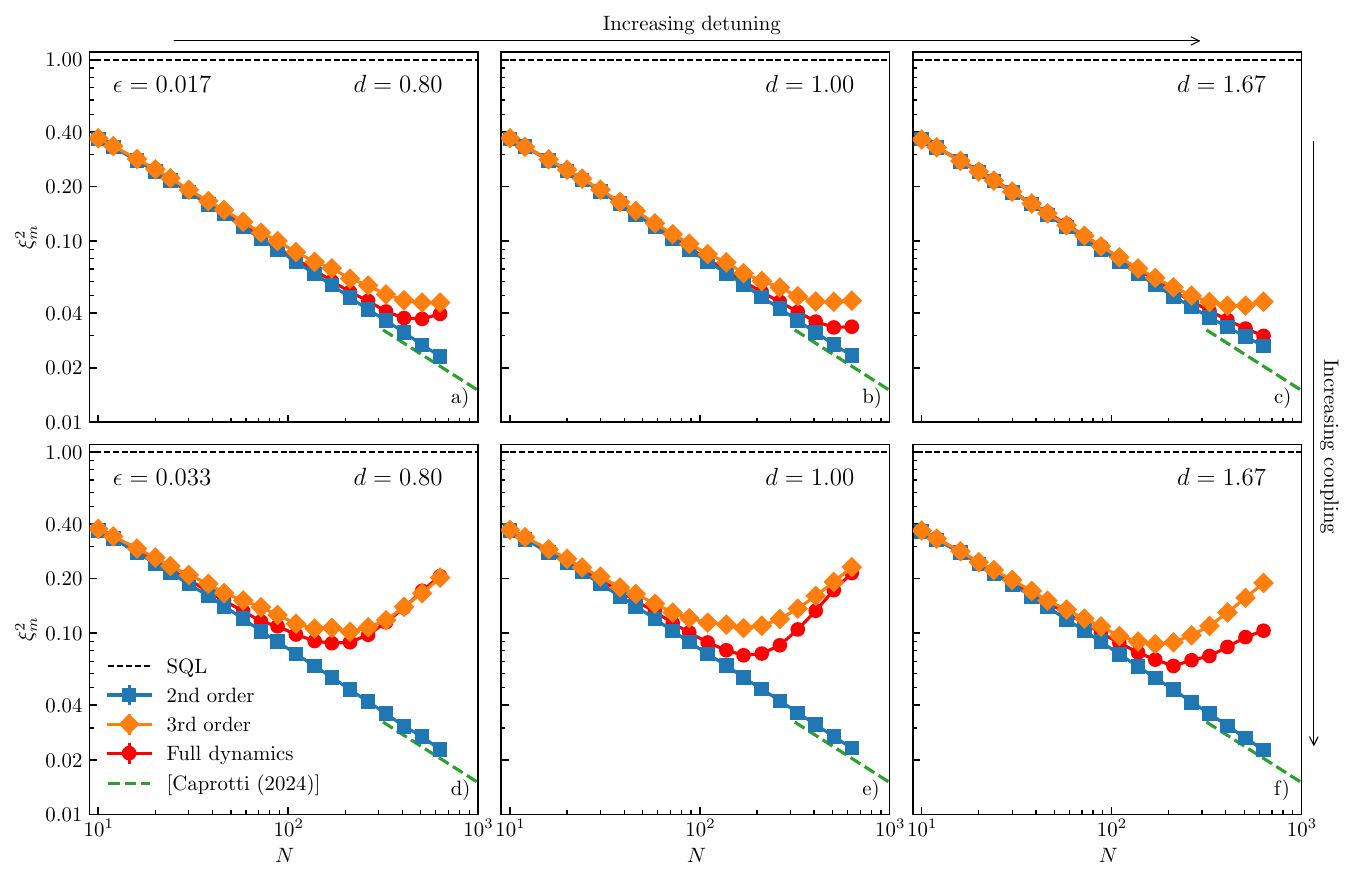}
 \caption{Values of the optimal (average) spin-squeezing parameter with increasing number of atoms $N$, for the conditional dynamics with coupling parameters and driving laser detunings arrayed as in Fig.~\ref{fig:compare_unconditional}, the optimal homodyne phase of Eq.~\eqref{eq:optimalphase}, and efficiency $\eta=1$. Circles: solution of the full Eq.~\eqref{eq:StochasticFullMasterEq}. Diamonds: solution of the third-order Eq.~\eqref{eq:StochasticRedMasterEq}. Squares: solution of Eq.~\eqref{eq:StochasticRedMasterEq} truncated to second order. Dashed line: second-order asymptotic result in the case $d=0$, Eq.~\eqref{eq:scalingcaprotti}. Horizontal dashed line: standard quantum limit. Statistical error bars are smaller than symbol sizes.}
\label{fig:compare_conditional_phiBarberena}
\end{figure*}
\subsection{QND conditional dynamics at finite detuning} 
Finally, in Fig.~\ref{fig:compare_conditional_phiBarberena}, we show the results of QuTiP simulations of the conditional dynamics where both QND measurements and Hamiltonian spin squeezing generation are active, namely at finite $d$, considering the same values of $\epsilon$ and $d$ as in Fig.~\ref{fig:compare_unconditional}. In these simulations, we fix the physical homodyne phase to the $d-$dependent value given by Eq.~\eqref{eq:optimalphase} (more details about this can be found in Sec.~\ref{sec:AtomCavity}), and consider the efficiency $\eta=1$. 
With increasing $\epsilon$ or decreasing $d$, the full simulations manifest the generation of less squeezing, analogously to the unconditional case, even though the relative changes are less marked. The simulations in the second row of Fig.~\ref{fig:compare_conditional_phiBarberena} clearly display an optimum in the spin-squeezing parameter at a certain number of atoms. Quite interestingly, differently from the case of $d=0$ of Fig.~\ref{fig:compare_conditional_d0}, here the third-order results reproduce this effect with reasonable accuracy, even though their dependence on $d$ appears to be much less marked than in the full simulations. 
We have also numerical evidence (not shown) that doing these conditional simulations with a non-optimal homodyne phase, such as $\phi=0$, yields both less spin squeezing and a worse comparison between the full and reduced dynamics.

\section{Adiabatic elimination for bipartite open quantum systems}
\label{sec:AdiabaticElBipartite}
In this section, we provide a brief yet comprehensive description of the adiabatic elimination method introduced in~\cite{7798963, Azouit_2017,10886784}, where second-order formulas for the solutions of the associated Sylvester equations are derived, and provide generic formulas for the third-order contributions, which we will apply to the considered atom-cavity system in Sec.~\ref{sec:AtomCavity}. Central to this reduction method based on central manifold theory is the presence of an invariant subspace corresponding to slow dynamics and characterized by the (right) eigenoperators of the Lindbladian whose eigenvalues have real part close to zero. Model reduction introduces two maps: one parameterizing the DOFs on the invariant manifold, $\mathcal{K}$, and the other describing the time evolution of the parameters, $\mathcal{L}_s$. In Ref.~\cite{Azouit_2017}, it is assumed that the fast-decaying subsystem exponentially converges to a unique steady state, i.e. that its Lindbladian has a simple zero eigenvalue and all other eigenvalues have strictly negative real part. 
The method developed there allows one, in the case of weak interaction, to systematically compute both $\mathcal{K}$ and $\mathcal{L}_s$ maps as asymptotic expansions with respect to the timescale separation between the slow and fast dynamics.

For a bipartite open quantum system let $\mathcal{H} = \mathcal{H}_A \otimes \mathcal{H}_B$ be the total Hilbert space, where $\mathcal{H}_A$ ($\mathcal{H}_B$) is the Hilbert space of the slow (fast) subsystem. The density matrix $\rho$ follows the Lindblad equation
\begin{equation}
 \label{eq:model}
 \frac{d}{dt} \rho = \mathcal{L} (\rho) = -i (\hat{H}_A \otimes \hat{I}_B)^\times(\rho)
 + \mathcal{I}_A \otimes \mathcal{L}_B(\rho) 
 -i \hat{H}_{\text{int}}^\times( \rho)\,,
\end{equation}
where $\hat{I}_M$ is the identity operator and $\mathcal{I}_M$ the identity superoperator, associated with the Hilbert space $\mathcal{H}_M$, $M =\{ A,\, B\}$. We are therefore restricting to a unitary dynamics for the slow subsystem. $\mathcal{L}_B$ is the Lindbladian acting only on the Hilbert space $\mathcal{H}_B$,
\begin{equation}
 \mathcal{L}_B(\bullet) = -i \hat{H}_B^\times(\bullet) + \mathcal{D}[\hat{L}](\bullet)\,,
\end{equation}
where the jump operator $\hat{L}=\sqrt{\kappa}\,\hat{c}$ is associated to the damping rate $\kappa$ which identifies the fast dynamics. As for the interaction term between subsystems $A$ and $B$ we take the form
\begin{equation}
 \hat{H}_{\text{int}}= g \hat{A}\otimes\hat{B}\,,
\end{equation}
with $\hat{A}$ and $\hat{B}$ operators on $\mathcal{H}_A$ and $\mathcal{H}_B$ respectively and $g$ a small positive coupling parameter, identifying the slow dynamics. As mentioned above, once we have chosen a parametrization $\rho_s$ on the invariant manifold, our aim is to determine the superoperator $\mathcal{L}_s$ describing the time evolution of $\rho_s$, namely $(d/dt) \rho_s = \mathcal{L}_s (\rho_s)$, and the full-dynamics solution map $\mathcal{K}$ between $\rho_s$ and the solution $\rho$ of Eq.~\eqref{eq:model}, namely $\rho = \mathcal{K} (\rho_s)$. This is achieved solving for the invariance condition 
\begin{equation}
 \mathcal{K} (\mathcal{L}_s (\rho_s)) = \mathcal{L} (\mathcal{K}(\rho_s))\,,
 \label{eq:AE_inv1}
\end{equation}
which can be done order by order in terms of the asymptotic expansions
\begin{equation}
 \mathcal{K} = \sum_{n = 0}^\infty \epsilon^n \mathcal{K}_n, \ \ \ \ \ \ \ \ \mathcal{L}_s = \sum_{n = 0}^\infty \epsilon^n \mathcal{L}_{s,n}\,,
 \label{eq:AE_asymexp}
\end{equation}
where the perturbative parameter is $\epsilon=g/\kappa \ll 1$. For $\epsilon = 0 $ the invariant manifold $\mathcal{S}_0$, on which all eigenvalues are purely imaginary, is spanned by all the density matrices $\rho = \rho_s \otimes \bar{\rho}_B$, with $\rho_s$ any density matrix on $\mathcal{H}_A$ and $\bar{\rho}_B$ the unique steady state of $\mathcal{L}_B$ in $\mathcal{H}_B$. The solution to the above-stated problem is given by
\begin{equation}\label{eq:zeroth}
 \mathcal{L}_{s,0}(\rho_s) = -i \hat{H}_A^\times(\rho_s), \quad\mathcal{K}_0(\rho_s) = \rho_s \otimes \bar{\rho}_B\,,
\end{equation}
with $\rho_s = \text{Tr}_B (\rho)$. 
For $\epsilon\neq 0$ the invariant submanifold $\mathcal{S}_\epsilon$ is not the same as $\mathcal{S}_0$ because of the correlations created by $\hat{H}_{\text{int}}$ between the two subsystems $A$ and $B$, so that we have to solve, order by order, the equations
\begin{multline}
\label{eq:invariance_nthorder}
 -i (\hat{H}_{A}\otimes \hat{I}_B)^\times(\epsilon \mathcal{K}_n(\rho_s)) + \mathcal{I}_A \otimes {\mathcal{L}}_{B}\left(\epsilon \mathcal{K}_n(\rho_s)\right)-i \hat{H}_{\text{int}}^\times(\mathcal{K}_{n-1}(\rho_s)) \\ = \sum_{m=0}^n\mathcal{K}_m\left(\epsilon {\mathcal{L}}_{s,n-m}(\rho_s)\right)\,,
\end{multline}
which can be re-expressed as Sylvester equations,
\begin{multline}
\label{eq:Sylvester_nthorder}
 -i (\hat{H}_{A}\otimes \hat{I}_B)^\times(\epsilon \mathcal{K}_n(\rho_s)) + \mathcal{I}_A \otimes {\mathcal{L}}_{B}\left(\epsilon \mathcal{K}_n(\rho_s)\right)-\mathcal{K}_n\left(\epsilon {\mathcal{L}}_{s,0}(\rho_s)\right) \\=i \hat{H}_{\text{int}}^\times( \mathcal{K}_{n-1}(\rho_s)) + \sum_{m=0}^{n-1}\mathcal{K}_m\left(\epsilon {\mathcal{L}}_{s,n-m}(\rho_s)\right)\,,
\end{multline}
Note that the solutions \eqref{eq:AE_asymexp} of Eqs.~\eqref{eq:Sylvester_nthorder} are not unique. For instance, for any solution $\mathcal{L}_s,\mathcal{K}$ and any fixed unitary $\hat{U}$, $\mathcal{L}'_s(\bullet) = \hat{U} \mathcal{L}_s(\hat{U}^\dagger \bullet \hat{U}) \hat{U}^\dagger$ and $\mathcal{K}'(\bullet) = \mathcal{K} (\hat{U}^\dagger \bullet \hat{U})$ are also solutions of Eq.~\eqref{eq:AE_inv1}. There is actually a gauge freedom associated with the DOFs parametrization on the invariant manifold; in~\cite{Azouit_2017} it has been conjectured that there is a gauge choice such that the reduced dynamics is governed by a Lindblad equation and the assignment of reduced DOFs is a trace-preserving completely-positive map (namely a Kraus map) up to any order of the asymptotic expansion. In the case of an invariant subspace associated with eigenvalues near zero of the total Lindbladian, this conjecture has been proven rigorously up to the second order, showing that the evolution equation admits a Lindblad form~\cite{Azouit_2017} and that there always exists a gauge choice ensuring the Kraus map assignment~\cite{azouit:tel-01743808}. In the case of an invariant subspace featuring large, almost purely imaginary eigenvalues, i.e. of a persistent unitary dynamics after decay, similar results have been proven under technical conditions on the fast unitary dynamics~\cite{Forni_AdiabaticEliminationMultiPartite_2018,forni9028922}. More recently, this conjecture has been challenged by computations at higher orders in~\cite{tokiedaPhysRevA.109.062206}. For the time being, we make the gauge choice
\begin{equation}
 \rho_s = \text{Tr}_B (\rho) = \text{Tr}_B (\mathcal{K}(\rho_s))\,,
\end{equation}
which implies 
\begin{equation}
 \text{Tr}_B (\mathcal{K}_1(\rho_s)) = \text{Tr}_B (\mathcal{K}_2(\rho_s)) = \text{Tr}_B (\mathcal{K}_3(\rho_s))=\ldots =0\,.
\end{equation}
We shall show in Sec.~\ref{sec:AtomCavity} that with this gauge choice we can straightforwardly argue for completely positive Lindbladian dynamics for the atom-cavity system after adiabatic elimination of the cavity up to the third order.

We further observe that $\text{Tr}_B ({\mathcal{L}}_{B}(\bullet))=0$ and $\text{Tr}_B ([\hat{H}_{A},\bullet])= [\hat{H}_{A},\text{Tr}_B(\bullet)]$, so that tracing Eq.~\eqref{eq:invariance_nthorder} one obtains
\begin{align}
\epsilon {\mathcal{L}}_{s,n}(\rho_s) &= -i \text{Tr}_B\left(\hat{H}_{\text{int}}^\times(\mathcal{K}_{n-1}(\rho_s))\right) = -i g\left[\hat{A}, \text{Tr}_B\left((\hat{I}_A \otimes \hat{B})\mathcal{K}_{n-1}(\rho_s)\right) \right]\,.
\end{align}
The right-hand side of Eq.~\eqref{eq:Sylvester_nthorder} is therefore known at each given order $n$. 
In particular one obtains~\cite{10886784},
\begin{equation}
\epsilon \mathcal{L}_{s,1}(\rho_s) = -i g \text{Tr}_B\left(\hat{B}\bar{\rho}_B\right)\left[\hat{A},\rho_s\right]\,,\label{eq:Lindbladian1orderfinal}
\end{equation}
for $n=1$, and 
\begin{align}
 \epsilon \mathcal{K}_1(\rho_s) &= -i g\intinf dt \, \left( \hat{A}^{\text{-}} (t) \rho_s \otimes e^{t{\mathcal{L}}_{B}} (\hat{B}_0 \bar{\rho}_B)  - \rho_s \hat{A}^{\text{-}} (t) \otimes e^{t{\mathcal{L}}_{B}} (\bar{\rho}_B \hat{B}_0 )\right)\,,
 \label{eq:Kraus1order}
\end{align}
where $\hat{A}^{\text{-}} (t) = e^{-i t\hat{H}_{A}^\times} (\hat{A})$, $\hat{B}_0 = \hat{B} - \text{Tr}_B\left(\hat{B}\bar{\rho}_B\right) \hat{I}_B$.
Throughout the paper, the limits of the formal time integrals are meant to be from $0$ to $\infty$.
One thus obtains the second-order term
\begin{equation}
\label{eq:Lindbaldian2orderfinal}
 \epsilon^2 {\mathcal{L}}_{s,2}(\rho_s) =-g^2 \intinf dt \,\left(\left[\hat{A}, \hat{A}^{\text{-}} (t) \rho_s c(t) \right] - \left[\hat{A}, \rho_s \hat{A}^{\text{-}} (t) \tilde{c}(t)\right]\right)\,,
\end{equation}
with
\begin{equation}
 c(t) = \text{Tr}_B\left(\hat{B} e^{t{\mathcal{L}}_{B}} (\hat{B}_0 \bar{\rho}_B)\right) = \text{Tr}_B\left(e^{t{\mathcal{L}}^{\ast}_{B}}(\hat{B}) \hat{B}_0 \bar{\rho}_B\right)\,,\label{eq:c}
 \end{equation}
 \begin{equation}
 \tilde{c}(t) = \text{Tr}_B\left(\hat{B}e^{t{\mathcal{L}}_{B}} (\bar{\rho}_B \hat{B}_0 )\right) = \text{Tr}_B\left(e^{t{\mathcal{L}}^{\ast}_{B}}(\hat{B})\bar{\rho}_B \hat{B}_0\right)\,,\label{eq:ctilde}
\end{equation}
where ${\mathcal{L}}^{\ast}_{B}(\bullet) = i [\hat{I}_A \otimes \hat{H}_{B},\bullet]+\hat{L}^\dagger\bullet\hat{L} -\{\hat{L}^\dagger \hat{L}, \bullet\}/2$ is the adjoint superoperator of ${\mathcal{L}}_{B}$.
Using this result, the right-hand side of Eq.~\eqref{eq:Sylvester_nthorder} for $n=2$ is known and we get the solution
\begin{align}
 \epsilon^2 &\mathcal{K}_2(\rho_s) \nonumber\\
 =&\, g^2\intinf dt' \intinf dt \Biggl(- \left[ \hat{A}^{\text{-}} (t'), \hat{A}^{\text{-}} (t+t') \rho_s \right] \otimes e^{t'\mathcal{L}_B}\left(\frac12 \left\{\hat{B}_0, e^{t\mathcal{L}_B} (\hat{B}_0 \bar{\rho}_B)\right\}-c(t)\bar{\rho}_B\right) \nonumber \Biggr.\\
 &- \left\{\hat{A}^{\text{-}} (t'), \hat{A}^{\text{-}} (t+t') \rho_s \right\} \otimes e^{t'\mathcal{L}_B}\left(\frac12 \left[\hat{B}_0, e^{t\mathcal{L}_B} (\hat{B}_0 \bar{\rho}_B)\right]\right)\nonumber\\
 &+ \left[ \hat{A}^{\text{-}} (t'), \rho_s \hat{A}^{\text{-}} (t+t') \right] \otimes e^{t'\mathcal{L}_B}\left(\frac12 \left\{\hat{B}_0, e^{t\mathcal{L}_B} ( \bar{\rho}_B \hat{B}_0)\right\}-\tilde{c}(t)\bar{\rho}_B\right) \nonumber\\
 &+ \left\{ \hat{A}^{\text{-}} (t'), \rho_s\hat{A}^{\text{-}} (t + t') \right\} \otimes e^{t'\mathcal{L}_B}\left(\frac12 \left[\hat{B}_0, e^{t\mathcal{L}_B} ( \bar{\rho}_B \hat{B}_0)\right]\right)\nonumber\\
 &-\text{Tr}_B\left(\hat{B}\bar{\rho}_B\right) \left[\hat{A}^{\text{-}}(t'),\hat{A}^{\text{-}} (t + t') \right]\rho_s \otimes e^{(t+t')\mathcal{L}_B} (\hat{B}_0 \bar{\rho}_B)\nonumber\\
 &\Biggl.+\text{Tr}_B\left(\hat{B}\bar{\rho}_B\right)\rho_s \left[\hat{A}^{\text{-}}(t'),\hat{A}^{\text{-}} (t + t')\right] \otimes e^{(t+t')\mathcal{L}_B} (\bar{\rho}_B \hat{B}_0 )\Biggr)\label{eq:Kraus2order}\,.
\end{align}
The final formula for the third-order reduced Lindbladian is
\begin{align}
 \epsilon^3& {\mathcal{L}}_{s,3}(\rho_s) = -i g\left[\hat{A}, \text{Tr}_B\left((\hat{I}_A \otimes \hat{B})\epsilon^2\mathcal{K}_2(\rho_s)\right) \right] \,, \nonumber\\
 =&\, i g^3 \intinf dt' \intinf dt\Biggl( \left[\hat{A},\left[ \hat{A}^{\text{-}} (t'), \hat{A}^{\text{-}} (t+t') \rho_s \right]\right] \left(d(t,t')-c(t)\text{Tr}_B\left(\hat{B}\bar{\rho}_B\right)\right) \Biggr.\nonumber\\
 &- \left[\hat{A},\left[ \hat{A}^{\text{-}} (t'), \rho_s \hat{A}^{\text{-}} (t+t') \right]\right] \left(\tilde{d}(t,t')-\tilde{c}(t)\text{Tr}_B\left(\hat{B}\bar{\rho}_B\right)\right) \nonumber\\
 &+ \left[\hat{A},\left\{ \hat{A}^{\text{-}} (t'), \hat{A}^{\text{-}} (t+t') \rho_s \right\}\right] f(t,t')- \left[\hat{A},\left\{ \hat{A}^{\text{-}} (t'), \rho_s\hat{A}^{\text{-}} (t + t') \right\}\right] \tilde{f}(t,t')\nonumber\\
 &+\text{Tr}_B\left(\hat{B}\bar{\rho}_B\right) \left[\hat{A},\left[\hat{A}^{\text{-}}(t'),\hat{A}^{\text{-}} (t + t') \right]\right]\rho_s c(t+t') \nonumber\\
 &\Biggl. -\text{Tr}_B\left(\hat{B}\bar{\rho}_B\right)\rho_s \left[\hat{A},\left[\hat{A}^{\text{-}}(t'),\hat{A}^{\text{-}} (t + t')\right]\right] \tilde{c}(t+t')\Biggr)\,, \label{eq:SuperOp3order}
\end{align}
where
\begin{equation}
 d(t,t') = \frac12\text{Tr}_B\left(e^{t{\mathcal{L}}^{\ast}_{B}} (\left\{\hat{B}(t'), \hat{B}_0\right\}) \hat{B}_0 \bar{\rho}_B\right) \,,\label{eq:d}
\end{equation}
\begin{equation}
 f(t,t') = \frac12 \text{Tr}_B\left(e^{t{\mathcal{L}}^{\ast}_{B}} (\left[\hat{B}(t'),\hat{B}_0\right]) \hat{B}_0 \bar{\rho}_B\right)\,,\label{eq:f}
 \end{equation}
\begin{equation}
 \tilde{d}(t,t') = \frac12\text{Tr}_B\left(e^{t{\mathcal{L}}^{\ast}_{B}} (\left\{\hat{B}(t'), \hat{B}_0\right\}) \bar{\rho}_B \hat{B}_0\right) \,,\label{eq:dtilde}
 \end{equation}
\begin{equation}
 \tilde{f}(t,t') = \frac12 \text{Tr}_B\left(e^{t{\mathcal{L}}^{\ast}_{B}} (\left[\hat{B}(t'),\hat{B}_0\right]) \bar{\rho}_B \hat{B}_0\right)\,.\label{eq:ftilde}
\end{equation}
We point out that the operators inside the traces in the coefficients \eqref{eq:c}, \eqref{eq:ctilde}, \eqref{eq:d}-\eqref{eq:ftilde} are time-evolved by the adjoint Lindblad operator, i.e. they are in the Heisenberg picture. We refer to App.~\ref{app:details} for the details of the derivation.
\section{The atom-cavity system}
\label{sec:AtomCavity}
The system of interest in this work is an ensemble of atoms trapped in a tight lattice and off-resonantly coupled to a high-finesse cavity. In this dispersive regime, only two internal atomic levels are considered, together with a single cavity mode, while the auxiliary excited atomic level is eliminated via a standard procedure under the condition that the atomic transition energy is the largest energy scale~\cite{Walls_QuantumOptics_2025}, thereby yielding Hamiltonian~\eqref{eq:hamiltonian}. The adaptation from the $\Lambda-$ to the $V-$level structure is straightforward~\cite{Caprotti_Analysisspinsqueezinggeneration_2024}. For the sake of simplicity, we neglect atomic dissipative processes mediated by cavity coupling~\cite{Barberena:2023ham}.

In this section, we explicitly evaluate the coefficients appearing in the formulae of the previous section, for the considered atom-cavity system. The first step is to identify the bipartite structure of the system, whose dynamics is given by Eq.~\eqref{eq:MEoriginal}. The cavity's dissipative dynamics is governed by the superoperator 
\begin{equation}
 {\mathcal{L}}_{B}(\bullet) = -i (\hat{I}_A \otimes \hat{H}_{B})^\times(\bullet)+\mathcal{D}[\sqrt{\kappa}\hat{b}](\bullet)\,,
\end{equation}
with
\begin{equation}
 \hat{H}_{B} =-\delta \hat{b}^\dagger \hat{b}\,,
\end{equation}
whereas the atoms have a trivial unitary dynamics in their rotating frame
\begin{equation}
\label{eq:atom_dynamics}
 \hat{H}_{A} = 0\,,
\end{equation}
which significantly simplifies the computations. 
The two subsystems interact through the Hamiltonian interaction
\begin{equation}
 \hat{H}_{\text{int}} = g \hat{S}_z\left(\hat{b}^\dagger \hat{b}+\alpha \hat{b}^\dagger + \alpha^\ast \hat{b}\right) 
 \equiv g \hat{A}\otimes\hat{B}\,.
\end{equation}
where $\hat{A} =\hat{S}_z $ and $\hat{B} = \hat{b}^\dagger \hat{b} + \alpha \hat{b}^\dagger + \alpha^\ast \hat{b}$. 
Since $\hat{A}=\hat{A}^{\text{-}} (t)= \hat{S}_z$ because of \eqref{eq:atom_dynamics}, all the expressions of the previous section are significantly simplified, and in particular Eqs.~\eqref{eq:Lindbaldian2orderfinal} and \eqref{eq:SuperOp3order} reduce to
\begin{align}
 \epsilon^2 {\mathcal{L}}_{s,2}(\rho_s) =& -g^2 \left(C\left[\hat{A}, \hat{A} \rho_s \right] - C^\ast\left[\hat{A}, \rho_s \hat{A} \right]\right)\,,\label{eq:Lindbaldian2ordersimple}\\
 \epsilon^3 {\mathcal{L}}_{s,3}(\rho_s) =& i g^3 \left(D\hat{S}_z\left[\hat{S}_z,\left[\hat{S}_z, \rho_s \right]\right] - D^\ast\left[\hat{S}_z,\left[\hat{S}_z, \rho_s\right]\right] \hat{S}_z \right.\nonumber\\ 
 & \left.+ F\hat{S}_z\left[\hat{S}_z,\left\{\hat{S}_z, \rho_s \right\}\right] 
 + F^\ast\left[\hat{S}_z,\left\{\hat{S}_z, \rho_s\right\}\right] \hat{S}_z \right)\,,\label{eq:SuperOp3ordersimple}
\end{align}
where
\begin{align}
 C=\intinf dt \,c(t)\,,\quad
D = \intinf dt' \intinf dt \,d(t,t')\,, \quad
F = \intinf dt' \intinf dt \,f(t,t')\label{eq:coeffs}\,,
\end{align}
noticing that
\begin{align}
 C^\ast =\intinf dt \,\tilde{c}(t)\,, \quad
 D^\ast = \intinf dt' \intinf dt \,\tilde{d}(t,t')\,, \quad
 F^\ast = -\intinf dt' \intinf dt \,\tilde{f}(t,t')\,. 
\end{align}
In Appendix~\ref{app:detailssuperoperator} we show that
\begin{equation}
\label{eq:MEfirst}
 \epsilon {\mathcal{L}}_{s,1}(\rho_s) =0\,,
\end{equation}
and we evaluate the coefficients in Eqs.~\eqref{eq:Lindbaldian2ordersimple}-\eqref{eq:SuperOp3ordersimple} as
\begin{align}
 C & = \frac{2 i n_0(d-i)}{\kappa(1+d^2)^2}\,,\label{eq:C}\\
 D & = \frac{n_0}{\kappa^2}\frac{1}{(1+d^2)^3} \left(3 -d^2+ i {d \left(d^2+5\right)}\right)\,,\label{eq:D}\\
 F & = \frac{n_0}{\kappa^2}\frac{1}{(1+d^2)^3}\left({ 1-3 d^2} + i {d \left(3 -d^2\right)}\right)\,.\label{eq:F}
\end{align}
As shown in Appendix~\ref{app:detailssuperoperator}, Eq.~\eqref{eq:Lindbaldian2ordersimple} can then be rewritten in the Lindblad form
\begin{equation}
\label{eq:MEsecond}
 \mathcal{L}_{s,2}(\rho_s) = \frac{4 n_0\kappa}{(1+d^2)^2}\left(-i\frac{d}{2}\left[\hat{S}_z^2,\rho_s\right]+\left(\hat{S}_z\rho_s\hat{S}_z-\frac12\left\{\hat{S}_z^2,\rho_s\right\}\right)\right)
 \,,
\end{equation}
implying Eqs.~\eqref{eq:MEsecondordert} and~\eqref{eq:chikappatilde}, whereas Eq.~\eqref{eq:SuperOp3ordersimple} becomes
\begin{align}
 \mathcal{L}_{s,3}(\rho_s) &= \frac{4 n_0 \kappa}{(1+d^2)^2}\left(i\frac{1-d^2}{1+d^2}\left[\hat{S}_z^3, \rho_s \right]-\frac{2d}{1+d^2}\left\{\hat{S}_z^3, \rho_s \right\}  +\frac{2(d+i)}{1+d^2}\hat{S}_z \rho_s \hat{S}_z^2+ \text{h.c.}\right)\,.
 \label{eq:NonLindblad3order}
\end{align}
Putting together all the terms, we get the expression of the master equation up to third order
\begin{align}
 \frac{d}{dt}\rho_s &=\mathcal{L}_s(\rho_s)=\epsilon {\mathcal{L}}_{s,1}(\rho_s)+\epsilon^2 {\mathcal{L}}_{s,2}(\rho_s)+\epsilon^3 {\mathcal{L}}_{s,3}(\rho_s) =
 -i \hat{H}_s^\times (\rho_s)+\tilde{\mathcal{D}}_s(\rho_s)\,,
\end{align}
where the third-order Hamiltonian is given by Eq.~\eqref{eq:Hamiltonian3order},
\begin{equation}
 \hat{H}_s = 
 \frac{4 n_0 \kappa}{(1 + d^2)^2}
 \left(
 \frac{d}{2}\epsilon^2 \hat{S}_z^2
 - \frac{1 - d^2}{1 + d^2}\epsilon^3 \hat{S}_z^3
 \right)\,, 
\end{equation}
and the third-order dissipator is 
\begin{multline}\label{eq:dissipator3}
 \tilde{\mathcal{D}}_s(\bullet)=\frac{4 n_0\kappa}{(1+d^2)^2}\left(\epsilon^2\hat{S}_z\bullet\hat{S}_z-\frac12\left\{\epsilon^2\hat{S}_z^2,\bullet\right\} \right.\\
 \left. + \frac{2(d + i)}{1+d^2} \epsilon^3\hat{S}_z \bullet \hat{S}_z^2+ \frac{2(d -i)}{1+d^2} \epsilon^3\hat{S}_z^2\bullet \hat{S}_z 
 - \frac{2d}{1+d^2} \left\{\epsilon^3\hat{S}_z^3, \bullet \right\}\right)\,.
\end{multline}
At first sight, it is not obvious that this dissipator corresponds to completely positive dynamics. On the one hand, one physically expects that starting with the original master equation, in which only a single dissipation channel is present, should lead to an effective master equation in which only one jump operator is present; on the other hand, it seems that two jump operators are present in Eq.~\eqref{eq:dissipator3}, namely $\hat{S}_z$ and $\hat{S}_z^2$. The two observations can be reconciled, provided that a suitable combination of $\hat{S}_z$ and $\hat{S}_z^2$ yields a single jump operator with a positive dissipation rate. This can be achieved by including the $O[\epsilon^4]$ contribution $16 n_0\kappa\epsilon^4{(1+d^2)^{-3}} \hat{S}_z^2 \bullet \hat{S}_z^2$, that guarantees that only a single eigenvalue $\tilde{\kappa}$ of the Lindbladian matrix is different from zero and real positive, yielding the diagonalized dissipator $\mathcal{D}[\hat{L}_s](\bullet)$ with the jump operator $\hat{L}_s$ and the eigenvalue $\tilde{\kappa}$ given by Eqs.~\eqref{eq:Jump3order} and \eqref{eq:chikappatilde}. 
The inclusion of this $O(\epsilon^4)$ term to recover a Lindblad-form dissipator with a single jump operator is justified by the structure of the original full dynamics, which features only one dissipation channel. It is therefore reasonable to expect that this structural property is preserved after adiabatic elimination \cite{tokiedaPhysRevA.109.062206}. We thus obtain the completely positive dynamics up to third order, given by the Markovian master equation in the Lindblad form in Eq.~\eqref{eq:MEtotalPositive}.

The single jump operator $\hat{L}_s$, defined from the reduced dissipator $\mathcal{D}[\hat{L}_s]$ up to a phase, provides a way of extending the second-order SME~\eqref{eq:Stochastic_2Order} to higher orders. We make the conjecture that the reduced SME has the form~\eqref{eq:StochasticRedMasterEq}, namely it includes a back-action term of type $\sqrt{\eta}\,\mathcal{H}[\hat{L}_s e^{-i\phi}](\rho_{s})\,dW_t$, associated to the photocurrent variation $\sqrt{\eta}\,\langle \hat{L}_s e^{-i\phi} + \hat{L}_s^\dagger e^{i\phi}\rangle\,dt + dW_t$. This is the standard form of a SME with a single jump operator~\cite{Albarelli2024}. While it reduces to the unconditional case of Eq.~\eqref{eq:MEtotalPositive} when the noise is averaged, it also posits a mapping between the single quantum trajectories of the full and the reduced cases. 

Since, differently from Refs.~\cite{DOHERTY_1999,Thomsen2002a,Thomsen2002}, we consider generic scaled detuning $d$, the reduced jump phase has to be suitably tuned. Ref.~\cite{Barberena:2023ham} proposed a mean-field determination of such phase, that we report here for completeness. We start from the full unconditional master equation with the unshifted modes and substitute $\hat{S}_z$ with its average value:
\begin{align}
 \frac{d\rho}{dt} = -i \left[-\delta \hat{a}^\dagger\hat{a}+g\langle\hat{S}_z\rangle \hat{a}^\dagger\hat{a}+\beta(\hat{a}^\dagger + \hat{a}),\rho\right] + \kappa\left(\hat{a}\rho\hat{a}^\dagger-\frac{1}{2}\left\{\hat{a}^\dagger\hat{a},\rho\right\}\right), 
\end{align}
The stationary value of the field is determined by a mean-field shift of the detuning $\delta\to \delta_{\text{MF}}=\delta-g\langle\hat{S}_z\rangle$, namely 
\begin{equation}
 \alpha \to \alpha_{\text{MF}}=\frac{-i\beta}{-i\delta_{\text{MF}} + \kappa/2 }\approx \alpha \left(1+\frac{i\,g\langle\hat{S}_z\rangle}{i\delta-\kappa/2}\right)\,.
\end{equation}
This implies that the stationary average value of the original jump operator $\sqrt{\kappa}\,\hat{b}$ at the mean-field level is 
\begin{equation}
 \sqrt{\kappa}\langle\hat{b}\rangle \approx \sqrt{\kappa}(\alpha_{\text{MF}}-\alpha) =\frac{i\,g\,\alpha\sqrt{\kappa}\langle\hat{S}_z\rangle}{i\delta-\kappa/2}=\sqrt{\tilde{\kappa}}\,e^{i\theta_s}\langle\hat{S}_z\rangle\,.
\end{equation}
The reduced jump phase $e^{i\theta_s} = (d - i)^2/(1 + d^2)$ depends only on the scaled detuning $d$. We argue that at higher orders in $\epsilon$ it will not get modified, since in the considered model only the combination $\epsilon \hat{S}_z$ can appear, as we already consistently derived in the dissipator. Finally, we hypothesize that the relation between the average values also holds at the operator level, and thus obtain Eq.~\eqref{eq:Jump3order} for the reduced jump operator, namely
\begin{equation}
 \hat{L}_s = \sqrt{\tilde{\kappa}}\,e^{i\theta_s}
 \left(\hat{S}_z + \frac{2(d - i)\epsilon}{1 + d^2}\,\hat{S}_z^2\right)\,.
\end{equation}
An extended discussion on the validity of this derivation will be given elsewhere~\cite{GiaccariConditional2026}, while here we verify this prescription \emph{a posteriori} from the simulation results of Sec.~\ref{sec:Results}.

We now discuss the role of the homodyne phase $\phi$. Since the leading-order term of the jump operator, $\hat{L}_s^\text{MF}=\sqrt{\tilde{\kappa}}e^{i\theta_s}\hat{S}_z$, is proportional to $\hat{S}_z$, we argue that maximizing its contribution to the photocurrent will induce more spin squeezing in the atomic system. Therefore we look for the homodyne phase such that
\begin{equation}\label{eq:SNRderivative}
 \frac{\partial \left\langle \hat{L}_s^\text{MF} e^{-i\phi}+\hat{L}_s^{\text{MF}\dagger} e^{i\phi}\right\rangle}{\partial \phi} = 0\,,
\end{equation}
which yields the optimal $\phi$
\begin{equation}\label{eq:optimalphase}
 \phi=\theta_s=-2\arctan{(1/d)}\,,
\end{equation}
consistent with the choice made in Ref.~\cite{Barberena:2023ham}. Going beyond the leading order in Eq.~\eqref{eq:SNRderivative} and determining the optimal $\phi$ using also the next-order term in the jump operator does not provide better values of the spin-squeezing parameter, because it would optimize the measurement of a combination of $\langle\hat{S}_z\rangle$ and $\langle\hat{S}_z^2\rangle$.

\section{Conclusions}
\label{sec:Conclusions}
In this work, we have reported a study of spin squeezing generation in a system of cold atoms on a lattice in a leaky cavity driven by an external laser detuned with respect to the cavity resonance frequency, both in the absence and in the presence of continuous homodyne measurement of the transmitted field. We have shown evidence for a loss of scalability with the number of atoms when cavity-atom coupling increases with respect to the dissipation or measurement rate, at odds with asymptotic scaling behaviors often discussed in the literature. This was proven by numerical simulation of the full atom-cavity system and related to a newly computed third-order contribution from adiabatic elimination. Such next-order correction to the well-established second-order dynamics is sufficient to qualitatively change the optimal spin-squeezing parameter behavior at large atom numbers, in agreement with the full simulations. The loss of scalability may impact future applications in quantum technologies, since it implies that there exists an optimal number of atoms for spin squeezing generation, given atomic and cavity parameters. A complete study of this effect as the parameters involved in the model vary is left for future work. In particular, numerical efficiency is quite sensitive to the number $N$ of probes and a convenient strategy to overcome this difficulty could be the generalized cumulant expansion approach developed by~\cite{doi:10.1143/JPSJ.17.1100, Plankensteiner2022quantumcumulantsjl}. We finally remark that a proper demonstration of the adiabatic elimination for the stochastic master equation describing the conditional dynamics is still missing in the literature, and we also plan to properly tackle this problem in a future work~\cite{GiaccariConditional2026}.

The data that support the findings of this study are openly available at the following URL/DOI: \url{https://doi.org/10.5281/zenodo.17865044}.

\acknowledgments
We acknowledge useful discussions with A. Riva and D. Barberena. 
We acknowledge funding from Italian Ministry of Research and Next Generation EU via the PRIN 2022 Project CONTRABASS (Contract No. 2022KB2JJM). 
We acknowledge the CINECA awards IsCb5-PSIOQUAS and IsCc7-SDOQS under the ISCRA initiative, for the availability of high-performance computing resources and support. 

\appendix
\section{Data analysis}\label{app:data}
\begin{figure}[ptb]
 \centering
 \includegraphics[width=0.8\textwidth]{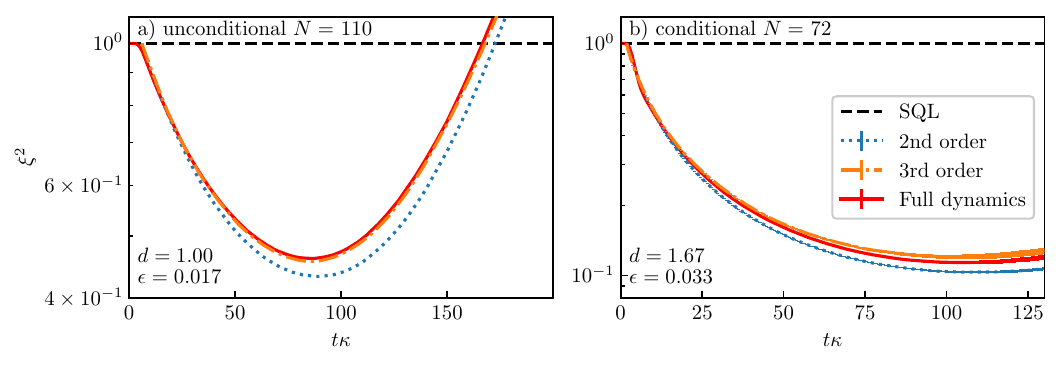}
 \caption{Time evolution of the spin-squeezing parameter. Panel a: unconditional dynamics for $N=110$, $\epsilon=g/\kappa = 0.017$, $d=2\delta/\kappa=1.0$. Panel b: Trajectory-averaged conditional dynamics for $N=72$, $\epsilon=g/\kappa = 0.033$, $d=2\delta/\kappa=1.67$, $\phi=\theta_s$ (Eq.~\eqref{eq:optimalphase}), with the error bars corresponding to the standard deviation of the mean. Solid lines: solution of the full equations. Dotted lines: solution of the second-order equations. Dot-dashed lines: solutions of the third-order equations. Horizontal dashed line: standard quantum limit. }
 \label{fig:dynamics_two}
\end{figure}

In Fig.~\ref{fig:dynamics_two}, we show two examples of the time evolution of the spin-squeezing parameter. On panel a, unconditional dynamics is simulated, using the \texttt{mcsolve} QuTiP routine. We solve the master equation of the full system, Eq.~\eqref{eq:MEoriginal}, the reduced second-order master equation, Eq.~\eqref{eq:MEsecondordert}, and the reduced third-order master equation, Eq.~\eqref{eq:MEtotalPositive}. 
On panel b, $M=1000$ conditional trajectories are simulated with the \texttt{ssesolve} QuTiP routine. We solve the SME of the full system, Eq.~\eqref{eq:StochasticFullMasterEq}, and the reduced SME, Eq.~\eqref{eq:StochasticRedMasterEq}, truncated to either second or third order. 
The spin-squeezing parameter is evaluated for each of the quantum trajectories (not shown), and its average over trajectories is shown in panel b, together with standard deviation error bars. Due to the computational burden of simulating many trajectories, the chosen maximum time of simulation is shorter in panel b than in panel a. In both panels, the reduced dynamics curves are manually time-shifted in order to compensate for the initial cavity-filling dynamics of the full case (see Ref.~\cite{Caprotti_Analysisspinsqueezinggeneration_2024}). Specific parameters of the simulated systems are reported in the caption, and other technical details are the same as discussed in Ref.~\cite{Caprotti_Analysisspinsqueezinggeneration_2024}. 
The minimum of the spin-squeezing parameter is then fitted from each curve. This yields the optimal $\xi_m^2$ at the considered $N$. Results for varying $N$ are finally collected to produce figures ~\ref{fig:compare_unconditional}-\ref{fig:compare_conditional_phiBarberena}.

\section{Solution to the invariance condition up to third order}\label{app:details}
We present here the details of the derivation of formulae \eqref{eq:Kraus1order}, \eqref{eq:Lindbaldian2orderfinal}, \eqref{eq:Kraus2order} and \eqref{eq:SuperOp3order} in Sec.~\ref{sec:AdiabaticElBipartite}.
We can spell out Eq.~\eqref{eq:Sylvester_nthorder} for the first few orders
\begin{equation}
\label{eq:invariance_I}
 -i (\hat{H}_{A}\otimes \hat{I}_B)^\times (\epsilon \mathcal{K}_1(\rho_s)) + \hat{I}_A \otimes \mathcal{L}_B\left(\epsilon \mathcal{K}_1(\rho_s)\right)-i \hat{H}_{\text{int}}^\times (\mathcal{K}_0(\rho_s)) 
 = \mathcal{K}_0\left(\epsilon \mathcal{L}_{s,1}(\rho_s)\right) + \epsilon \mathcal{K}_1\left(\mathcal{L}_{s,0}(\rho_s)\right)\,,
 \end{equation}

\begin{multline}
\label{eq:invariance_II}
 -i (\hat{H}_{A}\otimes \hat{I}_B)^\times(\epsilon^2 \mathcal{K}_2(\rho_s)) + \hat{I}_A \otimes \mathcal{L}_B\left(\epsilon^2 \mathcal{K}_2(\rho_s)\right)-i \hat{H}_{\text{int}}^\times(\epsilon \mathcal{K}_1(\rho_s)) \\
 = \mathcal{K}_0\left(\epsilon^2 \mathcal{L}_{s,2}(\rho_s)\right) + \epsilon \mathcal{K}_1\left(\epsilon\mathcal{L}_{s,1}(\rho_s)\right) + \epsilon^2 \mathcal{K}_2\left(\mathcal{L}_{s,0}(\rho_s)\right)\,,
 \end{multline}
\begin{multline}
 -i (\hat{H}_{A}\otimes \hat{I}_B)^\times(\epsilon^3 \mathcal{K}_3(\rho_s)) + \hat{I}_A \otimes \mathcal{L}_B\left(\epsilon^3 \mathcal{K}_3(\rho_s)\right)-i \hat{H}_{\text{int}}^\times(\epsilon^2 \mathcal{K}_2(\rho_s)) = \\
 = \mathcal{K}_0\left(\epsilon^3 \mathcal{L}_{s,3}(\rho_s)\right) + \epsilon \mathcal{K}_1\left(\epsilon^2\mathcal{L}_{s,2}(\rho_s)\right) + \epsilon^2 \mathcal{K}_2\left(\epsilon\mathcal{L}_{s,1}(\rho_s)\right) + \epsilon^3 \mathcal{K}_3\left(\mathcal{L}_{s,0}(\rho_s)\right)\,.
\end{multline}
Upon tracing out over the states of fast subspace, one finds
\begin{align}
 \epsilon \mathcal{L}_{s,1}(\rho_s) & = -i \text{Tr}_B\left(\hat{H}_{\text{int}}^\times(\mathcal{K}_0(\rho_s))\right)
 = -i g \text{Tr}_B\left(\hat{B}\bar{\rho}_B\right)\left[\hat{A},\rho_s\right]\,,\label{eq:LindbladOrder1}\\
 \epsilon \mathcal{L}_{s,2}(\rho_s) & = -i \text{Tr}_B\left(\hat{H}_{\text{int}}^\times (\mathcal{K}_1(\rho_s))\right)
 = -i g\left[\hat{A}, \text{Tr}_B\left((\hat{I}_A \otimes \hat{B})\mathcal{K}_1(\rho_s)\right) \right]\,,\label{eq:LindbladOrder2}\\
 \epsilon \mathcal{L}_{s,3}(\rho_s) & = -i \text{Tr}_B\left(\hat{H}_{\text{int}}^\times(\mathcal{K}_2(\rho_s))\right)
 = -i g\left[\hat{A}, \text{Tr}_B\left((\hat{I}_A \otimes \hat{B})\mathcal{K}_2(\rho_s)\right) \right]\,.
 \label{eq:LindbladOrder3}
\end{align}
We can then rewrite Eq.~\eqref{eq:invariance_I} as 
\begin{equation}
\label{eq:SylvesterKOrder1}
 -i (\hat{H}_{A}\otimes \hat{I}_B)^\times(\epsilon \mathcal{K}_1(\rho_s)) + \hat{I}_A \otimes \mathcal{L}_B\left(\epsilon \mathcal{K}_1(\rho_s)\right) - \epsilon \mathcal{K}_1\left(\mathcal{L}_{s,0}(\rho_s)\right) 
 = i \hat{H}_{\text{int}}^\times(\mathcal{K}_0(\rho_s)) + \mathcal{K}_0\left(\epsilon \mathcal{L}_{s,1}(\rho_s)\right)\,,
 \end{equation}
where the right-hand is known. 
Inserting Eq.~\eqref{eq:LindbladOrder1} in Eq.~\eqref{eq:SylvesterKOrder1}, we find
\begin{multline}
 -i (\hat{H}_{A}\otimes \hat{I}_B)^\times(\epsilon \mathcal{K}_1(\rho_s)) + \hat{I}_A \otimes \mathcal{L}_B\left(\epsilon \mathcal{K}_1(\rho_s)\right) +i \epsilon \mathcal{K}_1\left(\hat{H}_{A}^\times(\rho_s)\right) = \\
 = i g\left[\hat{A}\otimes\hat{B}, \rho_s \otimes \bar{\rho}_B\right] -i g \text{Tr}_B\left(\hat{B}\bar{\rho}_B\right)\left[\hat{A},\rho_s\otimes \bar{\rho}_B\right]= i g\left[\hat{A}\otimes\hat{B}_0, \rho_s \otimes \bar{\rho}_B\right]\,.
\end{multline}
The solution of this Sylvester equation for $\mathcal{K}_1(\rho_s)$ can be written as
\begin{multline}
 \epsilon \mathcal{K}_1(\rho_s) = - i g\intinf dt \, e^{t\left((-i \hat{H}_{A}\otimes \hat{I}_B)^\times + \hat{I}_A \otimes \mathcal{L}_B\right)}\left[\hat{A}\otimes\hat{B}_0, e^{ i t \hat{H}_{A}^\times}(\rho_s) \otimes \bar{\rho}_B\right]\\
 = -i g\intinf dt \, e^{-i t\hat{H}_{A}^\times} (\hat{A}) \rho_s \otimes e^{t\mathcal{L}_B} (\hat{B}_0 \bar{\rho}_B) - \rho_s e^{-i t\hat{H}_{A}^\times} (\hat{A})\otimes e^{t\mathcal{L}_B} (\bar{\rho}_B \hat{B}_0 )\,,
\end{multline}
which brings about Eq.~\eqref{eq:Kraus1order}. 
Inserting Eq.~\eqref{eq:Kraus1order} into Eq.~\eqref{eq:LindbladOrder2}, we get
\begin{equation}
\epsilon^2 \mathcal{L}_{s,2}(\rho_s) = g^2 \intinf dt \,\Biggl(
 \left[\hat{A},\rho_s \hat{A}^{\text{-}} (t) \text{Tr}_B\left(\hat{B}e^{t\mathcal{L}_B} (\bar{\rho}_B \hat{B}_0 )\right)\right] 
- \left[\hat{A},\hat{A}^{\text{-}} (t) \rho_s \text{Tr}_B\left(\hat{B} e^{t\mathcal{L}_B} (\hat{B}_0 \bar{\rho}_B)\right) \right]\Biggr)\,,
 \end{equation}
namely Eq.~\eqref{eq:Lindbaldian2orderfinal}. 
We can now rewrite Eq.~\eqref{eq:invariance_II} as
\begin{multline}
\label{eq:SylvesterKorder2}
 -i (\hat{H}_{A}\otimes \hat{I}_B)^\times(\epsilon^2 \mathcal{K}_2(\rho_s)) + \hat{I}_A \otimes \mathcal{L}_B\left(\epsilon^2 \mathcal{K}_2(\rho_s)\right) - \epsilon^2 \mathcal{K}_2\left(\mathcal{L}_{s,0}(\rho_s)\right)
 = \\
 = \mathcal{K}_0\left(\epsilon^2 \mathcal{L}_{s,2}(\rho_s)\right) + \epsilon \mathcal{K}_1\left(\epsilon\mathcal{L}_{s,1}(\rho_s)\right) + i \hat{H}_{\text{int}}^\times( \epsilon \mathcal{K}_1(\rho_s)) \,,
\end{multline}
where the terms on the right-hand side of the equation can now be explicitly written as
\begin{equation}
 \mathcal{K}_0\left(\epsilon^2 \mathcal{L}_{s,2}(\rho_s)\right) = -g^2 \intinf dt \,\left(\left[\hat{A}, \hat{A}^{\text{-}} (t) \rho_s \right] c(t) - \left[\hat{A},\rho_s \hat{A}^{\text{-}} (t) \right]\tilde{c}(t)\right)\otimes \bar{\rho}_B\,,
\end{equation}
\begin{multline}
 \epsilon \mathcal{K}_1\left(\epsilon\mathcal{L}_{s,1}(\rho_s)\right)\\
 = -g^2\,\text{Tr}_B\left(\hat{B}\bar{\rho}_B\right)\intinf dt
 \Biggl( \hat{A}^{\text{-}} (t) \left[\hat{A},\rho_s\right] \otimes e^{t\mathcal{L}_B} (\hat{B}_0 \bar{\rho}_B) 
-\left[\hat{A},\rho_s\right] \hat{A}^{\text{-}} (t) \otimes e^{t\mathcal{L}_B} (\bar{\rho}_B \hat{B}_0 )\Biggr) 
\end{multline}

\begin{align}
 i \hat{H}_{\text{int}}^\times(\epsilon \mathcal{K}_1(\rho_s)) &= g^2 \intinf dt \, \left[\hat{A} \otimes \hat{B}, \left(\hat{A}^{\text{-}} (t) \rho_s \otimes e^{t\mathcal{L}_B} (\hat{B}_0 \bar{\rho}_B) - \rho_s \hat{A}^{\text{-}} (t) \otimes e^{t\mathcal{L}_B} (\bar{\rho}_B \hat{B}_0 )\right)\right]\nonumber\\
 &= g^2 \intinf dt \, \Biggl(
 \left[\hat{A}, \hat{A}^{\text{-}} (t) \rho_s \right] \otimes \frac12 \left\{\hat{B}, e^{t\mathcal{L}_B} (\hat{B}_0 \bar{\rho}_B)\right\}
 + \left\{\hat{A}, \hat{A}^{\text{-}} (t) \rho_s \right\} \otimes \frac12 \left[\hat{B}, e^{t\mathcal{L}_B} (\hat{B}_0 \bar{\rho}_B)\right]\Biggr.\nonumber\\
 &\phantom{=}\Biggl.- \left[\hat{A}, \rho_s \hat{A}^{\text{-}} (t) \right] \otimes \frac12 \left\{\hat{B}, e^{t\mathcal{L}_B} ( \bar{\rho}_B \hat{B}_0)\right\}
 - \left\{\hat{A}, \rho_s\hat{A}^{\text{-}} (t) \right\} \otimes \frac12 \left[\hat{B}, e^{t\mathcal{L}_B} ( \bar{\rho}_B \hat{B}_0)\right]\Biggr)\,,
\end{align}
finally getting for the right-hand side of Eq.~\eqref{eq:SylvesterKorder2}:
\begin{align}
 \hat{X} =&\, i \hat{H}_{\text{int}}^\times( \epsilon \mathcal{K}_1(\rho_s)) +\epsilon \mathcal{K}_1\left(\epsilon\mathcal{L}_{s,1}(\rho_s)\right) + \mathcal{K}_0\left(\epsilon^2 \mathcal{L}_{s,2}(\rho_s)\right) \nonumber\\ 
 =&\, g^2\,\intinf dt \,
 \Biggl(\left[\hat{A}, \hat{A}^{\text{-}} (t) \rho_s \right] \otimes \left(\frac12 \left\{\hat{B}_0, e^{t\mathcal{L}_B} (\hat{B}_0 \bar{\rho}_B)\right\}-c(t)\bar{\rho}_B\right) \Biggr.\nonumber\\
 &- \left[\hat{A}, \rho_s \hat{A}^{\text{-}} (t) \right] \otimes \left(\frac12 \left\{\hat{B}_0, e^{t\mathcal{L}_B} ( \bar{\rho}_B \hat{B}_0)\right\}-\tilde{c}(t)\bar{\rho}_B\right) \nonumber\\
 &- \left\{\hat{A}, \rho_s\hat{A}^{\text{-}} (t) \right\} \otimes \frac12 \left[\hat{B}_0, e^{t\mathcal{L}_B} ( \bar{\rho}_B \hat{B}_0)\right] + \left\{\hat{A}, \hat{A}^{\text{-}} (t) \rho_s \right\} \otimes \frac12 \left[\hat{B}_0, e^{t\mathcal{L}_B} (\hat{B}_0 \bar{\rho}_B)\right]\nonumber\\
 &\Biggl.+\text{Tr}_B\left(\hat{B}\bar{\rho}_B\right) \left(\left[\hat{A},\hat{A}^{\text{-}} (t) \right]\rho_s \otimes e^{t\mathcal{L}_B} (\hat{B}_0 \bar{\rho}_B)
 -\rho_s \left[\hat{A},\hat{A}^{\text{-}} (t)\right] \otimes e^{t\mathcal{L}_B} (\bar{\rho}_B \hat{B}_0 )\right)\Biggr)\,.
\end{align}
We then have to solve
\begin{equation}
 -i (\hat{H}_{A}\otimes \hat{I}_B)^\times(\epsilon^2 \mathcal{K}_2(\rho_s)) + \hat{I}_A \otimes \mathcal{L}_B\left(\epsilon^2 \mathcal{K}_2(\rho_s)\right) +i \epsilon^2 \mathcal{K}_2\left( \hat{H}_{A}^\times(\rho_s)\right)
 = \hat{X}(\rho_s)\,
\end{equation}
finding the solution
\begin{equation}
 \epsilon^2 \mathcal{K}_2(\rho_s) = - \intinf dt' \,e^{t'\left((-i \hat{H}_{A}\otimes \hat{I}_B)^\times + \hat{I}_A \otimes \mathcal{L}_B\right)}\hat{X}(e^{ i t' \hat{H}_{A}^\times}(\rho_s))\,,
\end{equation}
which leads to Eq.~\eqref{eq:Kraus2order}. 
Inserting the latter into Eq.~\eqref{eq:LindbladOrder3}, we find
\begin{align}
 \epsilon^3 \mathcal{L}_{s,3}(\rho_s) = -i g\left[\hat{A}, \text{Tr}_B\left((\right.\right.&\left.\left.\hat{I}_A \otimes\hat{B})\epsilon^2\mathcal{K}_2(\rho_s)\right) \right]= \nonumber\\
 i g^3 \intinf dt' \intinf dt \, &\Biggl( \left[\hat{A},\left[\hat{A}^{\text{-}} (t'), \hat{A}^{\text{-}} (t+t') \rho_s \right]\right] \text{Tr}_B\left(\hat{B}e^{t'\mathcal{L}_B}\left(\frac12 \left\{\hat{B}_0, e^{t\mathcal{L}_B} (\hat{B}_0 \bar{\rho}_B)\right\}-c(t)\bar{\rho}_B\right)\right) \Biggr. \nonumber\\
 &+ \left[\hat{A},\left\{\hat{A}^{\text{-}} (t'), \hat{A}^{\text{-}} (t+t') \rho_s \right\}\right] \text{Tr}_B\left(\hat{B}e^{t'\mathcal{L}_B}\left(\frac12 \left[\hat{B}_0, e^{t\mathcal{L}_B} (\hat{B}_0 \bar{\rho}_B)\right]\right)\right)\nonumber\\
 &- \left[\hat{A},\left[\hat{A}^{\text{-}} (t'), \rho_s \hat{A}^{\text{-}} (t+t') \right]\right] \text{Tr}_B\left(\hat{B}e^{t'\mathcal{L}_B}\left(\frac12 \left\{\hat{B}_0, e^{t\mathcal{L}_B} ( \bar{\rho}_B \hat{B}_0)\right\}-\tilde{c}(t)\bar{\rho}_B\right)\right) \nonumber\\
 &- \left[\hat{A},\left\{ \hat{A}^{\text{-}} (t'), \rho_s\hat{A}^{\text{-}} (t + t') \right\}\right] \text{Tr}_B\left(\hat{B}e^{t'\mathcal{L}_B}\left(\frac12 \left[\hat{B}_0, e^{t\mathcal{L}_B} ( \bar{\rho}_B \hat{B}_0)\right]\right)\right)\nonumber\\
 &+\text{Tr}_B\left(\hat{B}\bar{\rho}_B\right) \left[\hat{A},\left[\hat{A}^{\text{-}}(t'),\hat{A}^{\text{-}} (t + t') \right]\right]\rho_s \text{Tr}_B\left(\hat{B}e^{(t+t')\mathcal{L}_B} (\hat{B}_0 \bar{\rho}_B)\right) \nonumber\\
 &\Biggl.-\,\text{Tr}_B\left(\hat{B}\bar{\rho}_B\right)\rho_s \left[\hat{A},\left[\hat{A}^{\text{-}}(t'),\hat{A}^{\text{-}} (t + t')\right]\right] \text{Tr}_B\left(\hat{B}e^{(t+t')\mathcal{L}_B} (\bar{\rho}_B \hat{B}_0 )\right)\Biggr)\,.
 \label{eq:SuperOp3orderImplicit}
 \end{align}
The traces showing up in this formula can be re-expressed in terms of a few coefficients dependent on the specific model under consideration. In particular,
\begin{align}
 \text{Tr}_B&\left(\hat{B}e^{t'\mathcal{L}_B}\left(\frac12 \left\{\hat{B}_0, e^{t\mathcal{L}_B} (\hat{B}_0 \bar{\rho}_B)\right\}-c(t)\bar{\rho}_B\right)\right)\nonumber\\
 &=\text{Tr}_B\left(\hat{B}e^{t'\mathcal{L}_B}\left(\frac12 \left\{\hat{B}_0, e^{t\mathcal{L}_B} (\hat{B}_0 \bar{\rho}_B)\right\}\right)\right)-\text{Tr}_B\left(\hat{B}e^{t'\mathcal{L}_B}\left(c(t)\bar{\rho}_B\right)\right)\nonumber\\
 &=\frac12\text{Tr}_B\left(\hat{B}(t') \left\{\hat{B}_0, e^{t\mathcal{L}_B} (\hat{B}_0 \bar{\rho}_B)\right\}\right)-c(t)\text{Tr}_B\left(\hat{B}\bar{\rho}_B\right)\nonumber\\
 &=\frac12\text{Tr}_B\left(\left\{\hat{B}(t'), \hat{B}_0\right\} e^{t\mathcal{L}_B} (\hat{B}_0 \bar{\rho}_B)\right)-c(t)\text{Tr}_B\left(\hat{B}\bar{\rho}_B\right)\nonumber\\
 &=\frac12\text{Tr}_B\left(e^{t\mathcal{L}^\ast_B} (\left\{\hat{B}(t'), \hat{B}_0\right\}) \hat{B}_0 \bar{\rho}_B\right)-c(t)\text{Tr}_B\left(\hat{B}\bar{\rho}_B\right)\nonumber\\
 &= d(t,t')-c(t)\text{Tr}_B\left(\hat{B}\bar{\rho}_B\right)\,,
\end{align}
and
\begin{multline}
 \text{Tr}_B\left(\hat{B}e^{t'\mathcal{L}_B}\left(\frac12 \left[\hat{B}_0, e^{t\mathcal{L}_B} (\hat{B}_0 \bar{\rho}_B)\right]\right)\right)=
 \frac12 \text{Tr}_B\left(\hat{B}(t')\left[\hat{B}_0, e^{t\mathcal{L}_B} (\hat{B}_0 \bar{\rho}_B)\right]\right) \\
 = \frac12 \text{Tr}_B\left(\left[\hat{B}(t'),\hat{B}_0\right] e^{t\mathcal{L}_B} (\hat{B}_0 \bar{\rho}_B)\right)= f(t,t')\,,
\end{multline}

\begin{equation}
 \text{Tr}_B\left(\hat{B}e^{t'\mathcal{L}_B}\left(\frac12 \left\{\hat{B}_0, e^{t\mathcal{L}_B} ( \bar{\rho}_B \hat{B}_0)\right\}-\tilde{c}(t)\bar{\rho}_B\right)\right)
 = \tilde{d}(t,t')-\tilde{c}(t)\text{Tr}_B\left(\hat{B}\bar{\rho}_B\right)\,,
\end{equation}

\begin{equation}
\text{Tr}_B\left(\hat{B}e^{t'\mathcal{L}_B}\left(\frac12 \left[\hat{B}_0, e^{t\mathcal{L}_B} ( \bar{\rho}_B \hat{B}_0)\right]\right)\right)
= \tilde{f}(t,t')\,,
\end{equation}
so that Eq.~\eqref{eq:SuperOp3orderImplicit} gives Eq.~\eqref{eq:SuperOp3order}.

\section{Third-order expansion of reduced Lindblad superoperator $\mathcal{L}_s$ for the atom-cavity system}
\label{app:detailssuperoperator}
In this Appendix, we give the details of the derivation of Eqs.~\eqref{eq:MEsecond} and \eqref{eq:NonLindblad3order}. 
We take advantage of the fact that the dynamics of the cavity subsystem is simple enough that the time evolution can be easily computed for any operator in the Heisenberg picture. 
We assume that the unique steady state of the cavity subsystem is described by $\bar{\rho}_B= \op{0}{0}_{B}$, where $\ket{0}_B$ is the driven cavity vacuum defined by $\hat{b}\ket{0}_B=0$, and therefore we have $\text{Tr}_B\left(\hat{B}\bar{\rho}_B\right) = 0$, so that $\epsilon {\mathcal{L}}_{s,1}(\rho_s) =0$. 
Given the cavity dynamics adjunct operator ${\mathcal{L}}^{\ast}_{B}(\bullet) = i (\hat{I}_A \otimes \hat{H}_{B})^\times(\bullet)+\kappa(\hat{b}^\dagger\bullet\hat{b} -\{\hat{b}^\dagger \hat{b}, \bullet\}/2)$, the first step is to show that the adjoint dynamics of operators expressed in terms of the cavity photon creation and destruction operators can be systematically derived. This guarantees the calculability of the model-dependent functions \eqref{eq:c}, \eqref{eq:ctilde}, \eqref{eq:d}-\eqref{eq:ftilde} entering formulae \eqref{eq:Kraus1order}, \eqref{eq:Lindbaldian2orderfinal}, \eqref{eq:Kraus2order} and \eqref{eq:SuperOp3order}. 
We can prove that in general the following holds for any ordered operator 
\begin{equation}
 \label{eq:TimeEvolutionOrderedOp} \mathcal{L}^\ast_B\left((\hat{b}^\dagger)^m (\hat{b})^n\right) =-\left(i\delta(m-n)+\frac{\kappa}{2}(m+n)\right)(\hat{b}^\dagger)^m (\hat{b})^n\,,
\end{equation}
so that
\begin{equation}
 e^{t\mathcal{L}^\ast_B}\left((\hat{b}^\dagger)^m (\hat{b})^n\right)= (\hat{b}^\dagger(t))^m (\hat{b}(t))^n\,.
\end{equation}
Notice however the same is not true for unordered operators since
$e^{t\mathcal{L}^\ast_B}\left([\hat{b},\hat{b}^\dagger]\right) = 1\neq \left[\hat{b}(t),\hat{b}^\dagger(t)\right]=e^{-\kappa t}$. 
Formula \eqref{eq:TimeEvolutionOrderedOp} is crucial in proving that 
\begin{align}
 \frac{d}{dt}\hat{B}(t)& = \mathcal{L}^\ast_B\left(e^{t\mathcal{L}^\ast_B}(\hat{B})\right)
 = e^{t\mathcal{L}^\ast_B}\left(\mathcal{L}^\ast_B(\hat{B})\right)= e^{t\mathcal{L}^\ast_B}\left(-\kappa \hat{b}^\dagger \hat{b}-\left(i \delta +\frac{\kappa}{2}\right)\alpha \hat{b}^\dagger + \left(i \delta -\frac{\kappa}{2}\right)\alpha^\ast \hat{b}\right)\nonumber\\
 & = -\kappa \hat{b}^\dagger (t)\hat{b}(t)-\left(i \delta +\frac{\kappa}{2}\right)\alpha \hat{b}^\dagger(t) + \left(i \delta -\frac{\kappa}{2}\right)\alpha^\ast \hat{b}(t)\,,
\end{align}

where $\hat{B}(t)= e^{t\mathcal{L}^\ast_B}\left(\hat{B}\right)$, 
$\hat{b}(t) = e^{\left(i \delta -\frac{\kappa}{2}\right)t}\hat{b}$ and $
 \hat{b}^\dagger (t) = e^{-\left(i \delta +\frac{\kappa}{2}\right)t}\hat{b}^\dagger$.
The solution is easily seen to be
\begin{equation}
 \hat{B}(t)= \hat{b}^\dagger (t)\hat{b}(t)+\alpha \hat{b}^\dagger(t) + \alpha^\ast \hat{b}(t)\,.
\end{equation}
One can then compute 
\begin{align}
 c(t) & = \text{Tr}_B\left(\hat{B}(t) \hat{B}_0 \bar{\rho}_B\right) = |\alpha|^2 e^{\left(i \delta -\frac{\kappa}{2}\right)t}\,,\nonumber\\
 \tilde{c}(t) & = \text{Tr}_B\left(\hat{B}(t)\bar{\rho}_B \hat{B}_0\right)=|\alpha|^2e^{-\left(i \delta +\frac{\kappa}{2}\right)t}\,,
\end{align}
which can be substituted into Eq.~\eqref{eq:coeffs} to get Eqs.~\eqref{eq:C}-\eqref{eq:F} straightforwardly.
Using formula \eqref{eq:TimeEvolutionOrderedOp} we can also compute the coefficients \eqref{eq:d}-\eqref{eq:ftilde}. 
In particular, we have
\begin{align}
 d(t,t') &= \frac12 |\alpha|^2 e^{\left(i \delta -\frac{\kappa}{2}\right)t}\left(e^{-\kappa t'} + e^{\left(i \delta -\frac{\kappa}{2}\right)t'}\right)\,,\\
 f(t,t') &= \frac12 |\alpha|^2 e^{\left(i \delta -\frac{\kappa}{2}\right)t}\left(-e^{-\kappa t'} +e^{\left(i \delta -\frac{\kappa}{2}\right)t'}\right)\,,\\
 \tilde{d}(t,t') & = \frac12 |\alpha|^2 e^{-\left(i \delta +\frac{\kappa}{2}\right)t}\left(e^{-\kappa t'} + e^{-\left(i \delta +\frac{\kappa}{2}\right)t'}\right) = {d}^\ast(t,t')\,,\\
 \tilde{f}(t,t') & = \frac12 |\alpha|^2 e^{-\left(i \delta +\frac{\kappa}{2}\right)t}\left(e^{-\kappa t'}-e^{-\left(i \delta +\frac{\kappa}{2}\right)t'}\right) = - {f}^\ast(t,t')\,.
\end{align}
One can readily infer that the above procedure could be extended to arbitrary order. We leave this for future work.

\vspace{0.5cm}
\providecommand{\newblock}{}

\end{document}